%% file: main.tex
\documentclass[conference]{IEEEtran}
\IEEEoverridecommandlockouts
\usepackage{cite}
\usepackage{amsmath,amssymb,amsfonts,amsthm}
\usepackage{algorithm}
\usepackage{algorithmic}
\usepackage{graphicx}
\usepackage{textcomp}
\usepackage{xcolor}
\usepackage{optidef}
\usepackage{pgfplots}
\usepackage{tikz}
\usepgfplotslibrary{fillbetween}
\usetikzlibrary{plotmarks, automata, positioning}
\usepackage[caption=false,font=footnotesize]{subfig}
\usepackage{url}

\input{macro.tex}

\def\BibTeX{{\rm B\kern-.05em{\sc i\kern-.025em b}\kern-.08em
    T\kern-.1667em\lower.7ex\hbox{E}\kern-.125emX}}
\begin{document}

\title{Password Strength Signaling: A Counter-Intuitive Defense Against Password Cracking
{\footnotesize \textsuperscript{}}
}

\author{\IEEEauthorblockN{Wenjie Bai}
\IEEEauthorblockA{\textit{Dept. of Computer Science} \\
\textit{Purdue University}\\
West Lafayette, USA \\
bai104@purdue.edu}
\and
\IEEEauthorblockN{Jeremiah Blocki}
\IEEEauthorblockA{\textit{Dept. of Computer Science} \\
\textit{Purdue University}\\
West Lafayette, USA \\
jblocki@purdue.edu}
\and
\IEEEauthorblockN{Ben Harsha}
\IEEEauthorblockA{\textit{Dept. of Computer Science} \\
\textit{Purdue University}\\
West Lafayette, USA \\
bharsha@purdue.edu}}

\maketitle

\begin{abstract}
\input{sections/abstract.tex}
\end{abstract}

\begin{IEEEkeywords}
Bayesian Persuasion, Password Authentication, Stackelberg Game
\end{IEEEkeywords}

\section{Introduction}\label{sec:intro}
\input{sections/intro.tex}

\section{Related Work}\label{sec:related}
\input{sections/related.tex}

\section{Preliminaries}\label{sec:background}
\input{sections/background.tex}

\section{Information Signaling and Password Storage}\label{sec:authmodel}
\input{sections/server.tex}

\section{Adversary Model}\label{sec:adversarymodel}
\input{sections/adversary.tex}

\section{Information Signaling as a Stackelberg Game}\label{sec:stackelberg}
\input{sections/game.tex}

\vspace{-0.05cm}
\section{Theoretical Example}
\vspace{-0.05cm}
\label{sec:theoretical_example}
\input{sections/theoretical_example.tex}

\section{Experimental Design}\label{sec:implementation}
\input{sections/implementation.tex}

\vspace{-0.1cm}
\section{Empirical Analysis}\label{sec:empirical}
\input{sections/experiments}

\vspace{-0.1cm}
\section{Conclusions}\label{sec:conclusions}
\input{sections/conclusion}

\section*{Acknowledgement}
This work was supported by NSF grant number 1755708 and Rolls-Royce through a Doctoral Fellowship.

\bibliographystyle{ieeetr}
\bibliography{bib/signal,bib/abbrev3,bib/crypto,bib/bounded-parallel-mhf,bib/jit,bib/extra}
\appendix

\input{sections/AppendixCompressedDistribution}

\section*{Extra Plots}
\input{figs/emprical_extra}
\input{figs/online_extra}
\end{document}

%% file: macro.tex
\definecolor{darkgreen}{rgb}{0.0,0.5,0.0}

\newcommand{\secref}[1]{Section \ref{#1}}

\newcommand{\sm}{\mathbf{S}}
\newcommand{\osm}{\mathbf{S}^*}
\newcommand{\getstr}[1]{\mathsf{getStrength}(#1)}
\newcommand{\getsig}[1]{\mathsf{getSignal}(#1)}
\newcommand{\ignore}[1]{} 

\newcommand{\tikzEmpirical}[0]{
  \pgfplotsset{
    ymin = 0,
    every axis/.append style={line width = 0.3pt},
    title style={align=center},
    xlabel={$v/C_{max}$},
    ylabel={Fraction of Cracked Passwords},
    log basis x={10},
    grid=major,
    cycle list = {{black}, {blue}, {green}, {blue!50!black, loosely dotted}, {magenta, dashed}, {cyan,dashed}, {red!50!black, loosely dotted}, {blue!50!black, loosely dashdotted}, {blue, densely dashdotted}, {blue, densely dashdotdotted}, {blue!50!black, loosely dotted}},
    legend style = {font=\tiny, at={(.01,.99)}, anchor=north west},
    legend entries = {no signal\\perfect knowledge signaling\\ imperfect knowledge signaling\\  improvement: {{\color{black}black- \color{blue} blue}}\\$\sm = \osm(10^5)$\\$\sm = \osm(10^6)$\\}
    }
}

\newcommand{\tikzMonte}[0]{
  \pgfplotsset{
    ymin =0,
    every axis/.append style={line width = 0.3pt},
    title style={align=center},
    xlabel={$v/C_{max}$},
    ylabel={Fraction of Cracked Passwords},
    log basis x={10},
    grid=major,
    cycle list = {{black}, {blue}, {green}, {blue!50!black, dotted}, {yellow, densely dashed},{blue, densely dashdotted}, {red!50!black, loosely dotted}, {blue!50!black, loosely dashdotted}, {blue, densely dashdotted}, {blue, densely dashdotdotted}, {blue!50!black, loosely dotted}},
    legend style = {font=\tiny, at={(.01,.99)}, anchor=north west},
    legend entries = {no signal\\ perfect knowledge signaling \\ imperfect knowledge signaling\\ improvement: {{\color{black}black- \color{blue} blue}}\\}
    }
}

\newcommand{\tikzUnlucky}[0]{
  \pgfplotsset{
    ymin =0,
    every axis/.append style={line width = 0.3pt},
    title style={align=center},
    xlabel={$v/C_{max}$},
    ylabel={Proportion of Unlucky Users},
    log basis x={10},
    grid=major,
    cycle list = {{blue, mark = o},{red, mark = triangle},{black, mark = +}},
    legend style = {font=\tiny, at={(.01,.99)}, anchor=north west},
    }
}

\newcommand{\tikzOnline}[0]{
  \pgfplotsset{
    ymin = 0,
    every axis/.append style={line width = 0.3pt},
    title style={align=center},
    xlabel={$v/C_{max}$},
    ylabel={Fraction of Cracked Passwords},
    log basis x={10},
    grid=major,
    cycle list = {{black}, {red}, {blue}, {blue!50!black, loosely dotted}, {cyan,dashed}, {magenta, dashed}, {red!50!black, loosely dotted}, {blue!50!black, loosely dashdotted}, {blue, densely dashdotted}, {blue, densely dashdotdotted}, {blue!50!black, loosely dotted}},
    legend style = {font=\tiny, at={(.01,.99)}, anchor=north west},
    legend entries = {no signal\\partition top 1k \\ partition top 10k\\  improvement: {{\color{black}black- \color{blue} blue}}\\$\sm = \osm(10^5)$\\$\sm = \osm(10^6)$\\}
    }
}

%% file: sections/abstract.tex
We introduce password strength signaling as a potential defense against password cracking.  Recent breaches have exposed billions of user passwords to the dangerous threat of offline password cracking attacks. An offline attacker can quickly check millions (or sometimes billions/trillions) of password guesses by comparing their hash value with the stolen hash from a breached authentication server. The attacker is limited only by the resources he is willing to invest. We explore the feasibility of applying ideas from Bayesian Persuasion to password authentication. Our key idea is to have the authentication server store a (noisy) signal about the strength of each user password for an offline attacker to find. Surprisingly, we show that the noise distribution for the signal can often be tuned so that a rational (profit-maximizing) attacker will crack {\em fewer} passwords. The signaling scheme exploits the fact that password cracking is not a zero-sum game i.e., the attacker's profit is given by the value of the cracked passwords {\em minus} the total guessing cost. Thus, a well-defined signaling strategy will encourage the attacker to reduce his guessing costs by cracking fewer passwords. We use an evolutionary algorithm to compute the optimal signaling scheme for the defender. We evaluate our mechanism on several password datasets and show that it can reduce the total number of cracked passwords by up to $12\%$ (resp. $5\%$) of all users in defending against offline (resp. online) attacks. While the results of our empirical analysis are positive we stress that we view the current solution as a proof-of-concept as there are important societal concerns that would need to be considered before adopting our password strength signaling solution. 

%% file: sections/intro.tex
   In the last decade, large scale data-breaches have exposed billions of user passwords to the dangerous threat of offline password cracking. An offline attacker who has obtained the (salted) cryptographic hash ($h_u=H(salt_u,pw_u)$) of a user $u$'s password $(pw_u)$  can attempt to crack the password by comparing this hash value with the hashes of likely password guesses i.e., by checking if $h_u^{\prime} = H(salt_u,pw')$ for each $pw'$. The attacker can check as many guesses as he wants offline  --- without interacting with the authentication server. The only limit is the resources that the attacker is willing to invest in trying to crack the password. A rational password cracker \cite{BlockiD16,SP:BloHarZho18} will choose the number of guesses that maximizes his utility.
   
   Password hashing serves as a last line of defense against an offline password attacker. A good password hash function $H$ should be moderately expensive to compute so that it becomes prohibitively expensive to check millions or billions of password guesses. However, we cannot make $H$ too expensive to compute as the honest authentication server needs to evaluate $H$ every time a user authenticates. In this paper, we explore a highly counter-intuitive\footnote{The propose may be less counter-intuitive to those familiar with prior work in the area of Bayesian Persuasion~\cite{Kamenica2011}.} defense against rational attackers which does not impact hashing costs: password strength signaling! In particular, we apply Bayesian Persuasion~\cite{Kamenica2011} to password authentication. Specifically, we propose to have the authentication server store a (noisy) signal $sig_u$ which is correlated with the strength of the user's password. 
   
   Traditionally, an authentication server stores the tuple  $(u,salt_u,h_u)$ for each user $u$ where $salt_u$ is a random salt value and $h_u = H(salt_u,pw_u)$ is the salted hash. We propose to have the authentication server instead store the tuple $(u,salt_u,sig_u, h_u)$, where the (noisy) signal $sig_u$ is sampled based on the strength of the user's password $pw_u$. The signal $sig_u$ is simply recorded for an offline attacker to find if the authentication server is breached. In fact, the authentication server never even uses $sig_u$ when the user $u$ authenticates\footnote{If a user $u$ attempts to login with password $pw'$ the authentication server will lookup $salt_u$ and $h_u$ and accept $pw'$ if and only if $h_u = H(salt_u,pw')$.}. The attacker will only use the signal $sig_u$ if it is beneficial --- at minimum the attacker could always choose to ignore the signal. 
   
It is natural, but incorrect, to imagine that password cracking is a zero-sum game i.e., the attacker's gain is directly proportional to the defender's loss.  In a zero-sum game there would be no benefit from information signaling \cite{AAMAS2016} e.g., in a zero-sum game like rock-paper-scissors there is no benefit to leaking information about your action. However, we stress that password cracking is {\em not} a zero-sum game. The defender's (the sender of strength signal) utility is inversely proportional to the fraction of user passwords that are cracked. By contrast, it is possible that the attacker's utility is marginal even when he cracks a password i.e., when guessing costs offset the reward. In particular, the attacker's utility is given by the (expected) value of all of the cracked passwords minus his (expected) guessing costs. Thus, it is possible that password strength signaling would persuade the attacker to crack fewer passwords to reduce guessing costs. Indeed, we show that the signal distribution can be tuned so that a rational (profit-maximizing) attacker will crack {\em fewer} passwords.

To provide some intuition of why information signaling might be beneficial, we give two examples.
\paragraph{Example 1} Suppose that we add a signal $sig_u=1$ to indicate that user $u$'s password $pw_u$ is uncrackable (e.g., the entropy of the password is over $60$-bits) and we add the signal $sig_u=0$ otherwise. In this case, the attacker will simply choose to ignore accounts with $sig_u=1$ to reduce his total guessing cost. However, the number of cracked user passwords stays unchanged.

\paragraph{Example 2} Suppose that we modify the signaling scheme above so that even when the user's password $pw_u$ is {\em not} deemed to be uncrackable we still signal $sig_u=1$ with probability $\epsilon$ and $sig_u=0$ otherwise. If the user's password is uncrackable we always signal $sig_u=1$. Assuming that $\epsilon$ is not too large a rational attacker might still choose to ignore any account with $sig_u=1$  i.e., the attacker's expected reward will decrease slightly, but the attacker's guessing costs will also be reduced. In this example, the fraction of cracked user passwords is reduced by up to $\epsilon$ i.e., any lucky user $u$ with $sig_u=1$ will not have their password cracked.

In this work, we explore the following questions: Can information signaling  be used to protect passwords against rational attackers? If so, how can we compute the optimal signaling strategy? 

\subsection{Contributions} We introduce password information signaling as a novel, counter-intuitive, defense against rational password attackers. We adapt a Stackelberg game-theoretic model of Blocki and Datta~\cite{BlockiD16} to characterize the behavior of a rational password adversary and the optimal signaling strategy for an authentication server (defender). We analyze the performance of password information signaling using several large password datasets: Bfield, Brazzers, Clixsense, CSDN, Neopets, 000webhost, RockYou, Yahoo!~\cite{SP:Bonneau12,NDSS:BloDatBon16}, and LinkedIn~\cite{LinkedInPwdCorpus}. We analyze our mechanism both in the idealistic setting, where the defender has perfect knowledge of the user password distribution $\mathcal{P}$ and the attacker's value $v$ for each cracked password, as well as in a more realistic setting where the defender only is given approximations of $\mathcal{P}$ and $v$. In our experiments, we analyze the fraction $x_{sig}(v)$ (resp. $x_{no-sig}(v)$) of passwords that a rational attacker would crack if the authentication server uses (resp. does not use) password information signaling.  We find that the reduction in the number of cracked passwords can be substantial e.g.,  $x_{no-sig}(v)-x_{sig}(v) \approx 8\%$ under empirical distribution and $13\%$ under Monte Carlo distribution. We also show that information signaling can be used to help deter online attacks when CAPTCHAs are used for throttling. 
 
An additional advantage of our information signaling method is that it is independent of the password hashing method and requires no additional hashing work. Implementation involves some determination of which signal to attach to a certain account, but beyond that, any future authentication attempts are handled exactly as they were before i.e. the signal information is ignored. 

We conclude by discussing several societal and ethical issues that would need to be addressed before password strength signaling is used. While password strength signaling decreases the total number of compromised accounts, there may be a few users whose accounts are cracked {\em because} they were assigned an ``unlucky" signal. One possible solution might be to allow users to opt-in (resp. opt-out). Another approach might try to constrain the solution space to ensure that there are no ``unlucky" users.

%% file: sections/related.tex

The human tendency to pick weaker passwords has been well documented e.g., \cite{SP:Bonneau12}. Convincing users to select stronger passwords is a difficult task~\cite{campbell2011impact,Komanduri2011,Shay2010,Stanton2005,Inglesant2010,Shay2014}. 
One line of research uses password strength meters to nudge users to select strong passwords~\cite{Komanduri2014,Ur2012,NDSS:CarMan14} though a common finding is that users were not persuaded to select a stronger password \cite{Ur2012,NDSS:CarMan14}. Another approach is to require users to follow stringent guidelines when they create their password. However it has been shown that these methods also suffer from usability issues \cite{Inglesant2010,NIST2014,Florencio2014lisa,Adams1999}, and in some cases can even lead to users selecting weaker passwords \cite{blockiPasswordComposition,Komanduri2011}. 

Offline password cracking attacks have been around for decades ~\cite{morris1979password}. There is a large body of research on password cracking techniques. State of the art cracking methods employ methods like Probabilistic Context-Free Grammars~\cite{SP:WAMG09,SP:KKMSVB12,NDSS:VerColTho14}, Markov models~\cite{NDSS:CasDurPer12,Castelluccia2013,SP:MYLL14,USENIX:USBCCKKMMS15}, and neural networks~\cite{USENIX:MUSKBCC16}. Further work~\cite{liu2019reasoning} has described methods of retrieving guessing numbers from commonly used tools like Hashcat~\cite{hashcat} and John the Ripper\cite{JohnTheRipper}.

A good password hashing algorithm should be moderately expensive so that it is prohibitively expensive for an offline attacker to check billions of password guesses. Password BCRYPT~\cite{provos1999bcrypt} or PBKDF2~\cite{kaliski2000pkcs} attempt to increase guessing this by iterating the hash function some number of times. However, Blocki et al. \cite{SP:BloHarZho18} argued that hash iteration cannot adequately deter an offline attacker due to the existence of sophisticated ASICs (e.g., Bitcoin miners) which can compute the underling hash function trillions of times per second. Instead, they advocate for the use of Memory Hard Functions (MHF) for password hashing.

MHFs at their core require some large amount of memory to compute in addition to longer computation times. Candidate MHFs include  SCRYPT~\cite{Per09}, Balloon hashing~\cite{AC:BonCorSch16}, and Argon2~\cite{Argon2} (the winner of the Password Hashing Competition\cite{PHC}). MHFs can be split into two distinct categories or modes of operation - data-independent MHFs (iMHFs) and data-dependent MHFs(dMHFs) (along with the hybrid idMHF, which runs in both modes). dMHFs like SCRYPT are maximally memory hard~\cite{EC:ACPRT17}, although they have the issue of possible side-channel attacks. Closely related to the notion of memory hardness is that of depth-robustness - a property of directed acyclic graphs (DAG). Alwen and Blocki showed that a depth robust DAG is both necessary~\cite{C:AlwBlo16} and sufficient~\cite{EC:AlwBloPie17} to construct a data-independent memory-hard function. Recent work has proposed candidate iMHF constructions that show resistance to currently-known attacks~\cite{blocki2019data}. Harsha and Blocki introduced a memory-hard KDF which accepts the input passwords as a stream so that the hashing algorithm can perform extra computation while the user is typing the password~\cite{harsha2018just}.

Blocki and Datta \cite{BlockiD16} used a Stackelberg game to model the behavior of a rational (profit-motivated) attacker against a cost-asymmetric secure hashing (CASH) scheme. However, the CASH mechanism is not easily integrated with modern memory-hard functions. By contrast, information signaling does not require any changes to the password hashing algorithm.

 Distributed hashing methods (e.g. ~\cite{USENIX:ECSJR15,CCS:CamLysNev12,USENIX:LESC17,USENIX:BJKS03}) offer a method to distribute storage and/or computation over multiple servers. Thus, an attacker who only breaches one server would not be able to mount an offline attack. Juels and Rivest proposed the inclusion of several false entries per user, with authentication attempts checked against an independent server to see if the correct entry was selected~\cite{CCS:JueRiv13}. These ``Honeyword" passwords serve as an alarm that an offline cracking attack is being attempted. Other methods of slowing down attackers include requiring some hard (for computers) problem to be solved after several failed authentication attempts (e.g. by using a CAPTCHA)~\cite{C:CanHalSte06,blockiGOTCHA,TCC:BloZho16}. An orthogonal line of research aims to protect users against online guessing attacks  \cite{tian2019stopguessing,DALock}.

A large body of research has focused on alternatives to text passwords. Alternatives have included one time passwords~\cite{ISC:FloHer08,pashalidis2004impostor, kuhn1998otpw}, challenge-response constructions~\cite{ESORICS:ChiVanBid07,jhawar2011make}, hardware tokens~\cite{authentication2003rsa, FC:ParKuoPer06}, and biometrics~\cite{ross2007template, daugman2009iris, aleksic2006audio}. While all of these offer possible alternatives to traditional passwords it has been noted that none of these strategies outperforms passwords in all areas~\cite{SP:BHVS12}. Furthermore, it has been noted that despite the shortcomings of passwords they remain the dominant method of authentication even today, and research should acknowledge this fact and seek to better understand traditional password use~\cite{herley2011research}. 
	

Password strength signaling is closely related to the literature on Bayesian Persuasion. Kamenica and Gentzkow~\cite{Kamenica2011} first introduced the notion of Bayesian Persuasion where a person (sender) chooses a signal to reveal to a receiver in an attempt to convince the receiver to take an action that positively impacts the welfare of both parties. Alonso and Camara \cite{alonso2016} studied the (sufficient) conditions under which a sender can benefit from persuasion. Dughmi et. al ~\cite{Dughmi2016} and Hoefer et. al \cite{hoefer2020} study Bayesian Persuasion from an algorithmic standpoint in different contexts. There are a few prior results applying Bayesian Persuasion in security contexts, e.g., patrols \cite{AAAI2011}, honeypots\cite{rabinovich2015}, with the sender (resp. receiver) playing the roles of defender (resp. attacker). To the best of our knowledge Bayesian Persuasion has never been applied in the context of password authentication.  In the most general case it is computationally intractable to compute the sender's optimal strategy\cite{Dughmi2016}. Most prior works use linear programming to find (or approximate) the sender's optimal signaling strategy. We stress that there are several unique challenges in the context of password authentication: (1) the action space of the receiver (attacker) is exponential in the size of (the support of) the password distribution, and (2) the sender's objective function is non-linear.

%% file: sections/background.tex
\subsection{Password Representation}
We use $\mathbb{P}$ to denote the set of all passwords that a user might select and we use $\mathcal{P}$ to denote a distribution over user selected passwords i.e., a new user will select the password $pw \in \mathbb{P}$ with probability $\Pr_{x \sim \mathcal{P}}[x=pw]$ --- we typically write $\Pr[pw]$ for notational simplicity.

\subsubsection{Password Datasets}
Given a set of $N$ users $\mathcal{U}=\{u_1,\ldots, u_N\}$ the corresponding password dataset $ D_u$ is given by the multiset $D_u = \{pw_{u_1},\ldots,pw_{u_N}\}$ where $pw_{u_i}$ denotes the password selected by user $u_i$. Fixing a password dataset $D$ we let $f_i$ denote the number of users who selected the $i$th most popular password in the dataset. We note that  that  $f_1 \geq f_2 \geq \ldots$ and that $\sum_i f_i = N$ gives the total number $N$ of users in the original dataset. 

\subsubsection{Empirical Password Distribution} Viewing our dataset $D$ as $N$ independent samples from the (unknown) distribution $\mathcal{P}$, we use $f_i/N$ as an empirical estimate of the probability of the $ith$ most common password $pw_i$ and $D_f = (f_1,f_2,\ldots)$ as the corresponding frequency list. In addition, $\mathcal{D}_e$ is used to denoted the corresponding empirical distribution i.e., $\Pr_{x \sim \mathcal{D}_e}[x=pw_i] = f_i/N$. Because the real distribution $\mathcal{P}$ is unknown we will typically work with the empirical distribution $\mathcal{D}_e$. We remark that when $f_i \gg 1$ the empirical estimate will be close to the actual distribution i.e., $\Pr[pw_i] \approx f_i/N$, but when $f_i$ is small the empirical estimate will likely diverge from the true probability value. 
Thus, while the empirical distribution is useful to analyze the performance of information signaling,  when the password value $v$ is small this analysis will be  less accurate for larger values of $v$ i.e., once the rational attacker has incentive to start cracking passwords with lower frequency.

\subsubsection{Monte Carlo Password Distribution} Following \cite{DAHash} we also use the Monte Carlo Password Distribution $\mathcal{D}_m$ to evaluate the performance of our password signaling mechanism when $v$ is large. The Monte Carlo distributions is derived by subsampling passwords from our dataset $D$, generating guessing numbers from state of the art password cracking models, and  fitting a distribution to the resulting guessing curve.    See more details in section \ref{sec:implementation}.

\subsubsection{Password Equivalence Set}  It is often convenient to group passwords having (approxmiately) equal probability into an \emph{equivalence set} $es$. Suppose there are $N^{\prime}$ equivalence sets, we typically have $N' \ll N$. Thus, an algorithm whose running time scales with $n'$ is much faster than an algorithm whose running time scales with $N$, see Appendix \ref{appendix:es}.


\subsection{Differential Privacy and Count Sketches}
As part of our information signaling, we need a way for the authentication server to estimate the strength of each user's passwords. We propose to do this  with a (differentially private) Count-Sketch data structure, which allows us to approximately determine how many users have selected each particular password. As a side-benefit the authentication server could also use the Count-Sketch data structure to identify/ban overly popular passwords~\cite{schechter2010popularity} and to defend against online guessing attacks~\cite{DALock,tian2019stopguessing}. We first introduce the notion of differential privacy.

\subsubsection{$\epsilon$-Differential Privacy}
$\epsilon$-Differential Privacy~\cite{TCC:DMNS06} is a mechanism that provides strong information-theoretic privacy guarantees for all individuals in a dataset.  Formally, an algorithm $\mathcal{A}$ preserves $\epsilon$-differential privacy iff for all datasets $D$ and $D'$ that differ by only one element and all subsets $S$ of $\text{Range}(\mathcal{A})$:
$$
\Pr\left[\mathcal{A}(D) \in S \right] \leq \mathrm{e}^\epsilon \Pr\left[\mathcal{A}(D') \in S\right].
$$
In our context, we can think of $D$ (resp. $D'$) as a password dataset which does (resp. does not) include our user $u$'s password $pw_u$ and we can think of $\mathcal{A}$ as a randomized algorithm that outputs a noisy count-sketch algorithm. Intuitively, differential privacy guarantees that an attacker cannot even tell if $pw_u$ was included when the count-sketch was generated. In particular, (up to a small multiplicative factor $e^{\epsilon}$) the attacker cannot tell the difference between $\mathcal{A}(D)$ and $\mathcal{A}(D')$ the count-sketch we sample when $pw_u$ was (resp. was not) included. Thus, whatever the attacker hopes to know about $u$'s from  $\mathcal{A}(D)$ the attacker could have learned from $\mathcal{A}(D')$.

\subsubsection{Count-sketch}
A count sketch over some domain $E$ is a probabilistic data structure that stores some information about the frequency of items seen in a stream of data --- in our password context we will use the domain $E = \mathbb{P}$. A count-sketch functions as a table $T$ with width $w_s$ columns and depth $d_s$ rows. Initially, $T[i,j]=0$ for all $i \leq w_s$ and $j \leq d_s$.  Each row is associated with a hash function $H_i: \mathbb{P} \rightarrow [w_s]$, with each of the hash functions used in the sketch being pairwise independent. 

To insert an element $pw \in \mathbb{P}$ into the count sketch we update $T[i,H_i(pw)]\leftarrow T[i,H_i(pw)]+1$ for each $i \leq d_s$ \footnote{In some instantiations of count sketch we would instead set $T[i,H_i(pw)]\leftarrow T[i,H_i(pw)]+G_i(pw)$ where the hash function $G_i: \mathbb{P} \rightarrow \{-1,1\}$}. To estimate the frequency of $pw$ we would output $f\left( T[1,H_1(pw)],\ldots , T[d_s,H_{d_s}(pw)]\right)$ for some function $f:\mathbb{N}^{d_s} \rightarrow \mathbb{N}$. In our experiments we instantiate a Count-Mean-Min Sketch where $f = \mathtt{median}\left\{T[i,H_i(pw)] - \frac{\#total - T[i,H_i(pw)]}{d_w-1}:i=1,\ldots,d_s\right\}$ ($\#total$ is the total number of elements being inserted) so that bias is subtracted from overall estimate.  Other options are available too, e.g., $f = \mathtt{min}$ (Count-Min), $f = \mathtt{mean}$ (Count-Mean-Sketch) and $f=\mathtt{median}$ (Count-Median) \footnote{Count-Median Sketch uses a different insersion method}. 

Oserve that adding a password only alters the value of $T[i,j]$ at $d_s$ locations. Thus, to preserve $\epsilon$-differential privacy we can initialize each cell $T[i,j]$ by adding Laplace noise with scaling parameter $d_s/\epsilon$ ~\cite{cormode2012differentially}.

\subsection{Other Notation} Given a permutation $\pi$ over all allowable passwords $\mathbb{P}$ we let $\lambda(\pi,B) := \sum_{i=1}^B \Pr\left[pw_i^{\pi}\right]$ denote the probability that a randomly sampled password $pw \in \mathbb{P}$ would be cracked by an attacker who checks the first $B$ guesses according to the order $\pi$ --- here $pw_i^{\pi}$ is the $i$th password in the sequence $\pi$.  Given an randomized algorithm $\mathcal{A}$ and a random string $r$ we use $y \leftarrow \mathcal{A}(x;r)$ to denote the output when we run $\mathcal{A}$ with input $x$ fixing the outcome of the random coins to be $r$. We use $y \overset{\$}{\leftarrow} \mathcal{A}(x)$ to denote a random sample drawn by  sampling the random coins $r$ uniformly at random. Given a randomized (signaling) algorithm $\mathcal{A}:\mathbb{P} \rightarrow [0, b-1]$ (where $b$ is the total number of signals) we define the conditional probability $\Pr[pw~|~y]:= \Pr_{x \sim \mathcal{P}, r}[x=pw~|~y=\mathcal{A}(pw)]$ and \[ \lambda(\pi, B; y) := \sum_{i=1}^B \Pr[pw_i^{\pi}~|~y] \ . \]
We remark that $\Pr[pw~|~y]$ can be evaluated using Bayes Law given knowledge of the signaling algorithm $\mathcal{A}(x)$.




%% file: sections/server.tex
In this section, we overview our basic signaling mechanism deferring until later how to optimally tune the parameters of the mechanism to minimize the number of cracked passwords.

\subsection{Account Creation and Signaling}
When users create their accounts they provide a user name $u$ and password $pw_u$. First, the server runs the canonical password storage procedure---randomly selecting a salt value $salt_u$ and calculating the hash value $h_u=H(salt_u,pw_u)$. Next, the server calculates the (estimated) strength  $str_u \leftarrow \mathsf{getStrength}(pw_u)$ of password $pw_u$ and samples the signal  $sig_u  \overset{\$}{\leftarrow} \mathsf{getSignal}(st_u)$. Finally, the server stores the tuple $(u, salt_u, sig_u, h_u)$ --- later if the user $u$ attempts to login with a password $pw'$ the authentication server will accept $pw'$ if and only if $h_u = H(salt_u,pw')$. The account creation process is formally presented in Algorithm \ref{algo:account}. 

\begin{algorithm}[h]
\caption{Signaling during Account Creation}
\label{algo:account}
\begin{algorithmic}[1]
\REQUIRE{$u$, $pw_u$, $L$, $d$}
\STATE $salt_u \overset{\$}{\leftarrow} \{0,1\}^L$
\STATE $h_u \gets H(salt_u, pw_u)$
\STATE $str_u \gets \mathsf{getStrength}(pw_u)$
\STATE $sig_u \overset{\$}{\leftarrow} \mathsf{getSignal}(str_u)$
\STATE $\mathsf{StoreRecord}(u,salt_u, sig_u, h_u)$
\end{algorithmic}
\end{algorithm}

A traditional password hashing solution would simply store the tuple $(u, salt_u, h_u)$ i.e., excluding the signal $sig_u$. Our mechanism requires two additionally subroutines $\getstr$ and $\getsig$ to generate this signal. The first algorithm is deterministic. It takes the user's password $pw_u$ as input and outputs $str_u$ --- (an estimate of) the password strength. The second randomized algorithm takes the (estimated) strength parameter $str_u$ and outputs a signal $sig_u$. The whole signaling algorithm is the composition of these two subroutines i.e., $\mathcal{A} = \getsig{\getstr{pw}}$. We use $s_{i,j}$ to denote the probability of observing the signal $sig_u = j$ given that the estimated strength level was $str_u=i$. Thus, $\getsig$ can be encoded using a signaling matrix $\mathbf{S}$ of dimension $a\times b$, i.e.,
\begin{equation*}
	\begin{bmatrix}
		s_{0,0}&s_{0,1}&\cdots&s_{0,b-1}\\
		s_{1,0}&s_{1,1}&\cdots&s_{1,b-1}\\
		\vdots&\vdots&\ddots&\vdots\\
		s_{a-1,0}&s_{a-1,1}&\cdots&s_{a-1,b-1}\\
	\end{bmatrix},	
\end{equation*}
where $a$ is the number of strength levels that passwords can be labeled, $b$ is the number of signals the server can generate and $\mathbf{S}[i,j] = s_{i,j}$.

We remark that for some signaling matrices (e.g., if $\sm[i, 0] = 1$ for all $i$ \footnote{The index of matrix elements start from 0}) then the actual signal $sig_u$ is {\em uncorrelated} with the password $pw_u$. In this case our mechanism is equivalent to the traditional (salted) password storage mechanism where $\getsig$ is replaced with a constant/null function. $\getstr$ is password strength oracle that outputs the actual/estimated strength of a password. We discuss ways that $\getstr$ could be implemented in \secref{sec:implementation}. For now, we omit the implementation details of strength oracle $\getstr$ for sake of readability.

\subsection{Generating Signals} 
We use $[a]=0,1, \ldots,a-1$ (resp. $[b] = 0,1, \ldots, b-1$) to denote the range of $\getstr$ (resp. $\getsig$). For example, if $[a]=\{0, 1, 2\}$ then $0$ would correspond to weak passwords, $2$ would correspond to strong passwords and $1$ would correspond to medium strength passwords. To generate signal for $pw_u$, the server first invokes subroutine $\getstr{pw_u}$ to get strength level $str_u = i \in [a]$ of $pw_u$, then signals $sig_u = j \in [b]$ with probability $\Pr[\getsig{pw_u}=j~|~\getstr{pw_u}=i] = \mathbf{S}[i,j] = s_{i,j}$.

\paragraph{Bayesian Update} An attacker who breaks into the authentication server will be able to observe  the signal $sig_u$ and $\sm$. After observing the signal $sig_u=y$ and $\sm$ the attacker can perform a Bayesian update. In particular, given any password $pw \in \mathbb{P}$ with strength $i = \getstr{pw}$ we have
\begin{equation}
\begin{split}
&\Pr\left[pw \;\middle\vert\;  y\right] \\
&=\frac{\Pr[pw] \mathbf{S}[i,y]}{\sum_{pw' \in \mathbb{P}}\Pr\left[\mathsf{getSignal}\left(\mathsf{getStrength}(pw')\right)\right]\cdot \Pr\left[pw'\right]} \\
&= \frac{\Pr[pw] \mathbf{S}[i,y]}{\sum_{i' \in [a]} \Pr_{pw' \sim \mathcal{P}} [\mathsf{getStrength}(pw')=i'] \cdot \mathbf{S}[i',y]}\\
\end{split}
\end{equation} 
If the attacker knew the original password distribution $\mathcal{P}$ then s/he can update posterior distribution $\mathcal{P}_y$ with $\Pr_{x \sim \mathcal{P}_y}\left[x=pw\right] :=\Pr\left[pw \;\middle\vert\;  y\right]$. We extend our notation, let $\lambda(\pi,B;y) = \sum_{i=1}^B \Pr\left[pw_i^{\pi} \;\middle\vert\;  y\right]$ where $pw_i^{\pi}$ is the $i$th password in the ordering $\pi$. Intuitively,   $\lambda(\pi,B;y) $ is the conditional probability of cracking the user's password by checking the first $B$ guesses in permutation $\pi$.

\subsection{Delayed Signaling}
In some instances, the authentication server might implement the password strength oracle $\getstr$ by training a (differentially private) Count-Sketch based on the user-selected passwords $pw_u \sim \mathcal{P}$. In this case, the strength estimation will not be accurate until a larger number $N$ of users have registered. In this case, the authentication server may want to delay signaling until after the Count-Sketch has been initialized. In particular, the authentication server will store the tuple $(u,salt_u, sig_u = \bot, h_u)$. During the next (successful) login with the password $pw_u$ the server can update $sig_u = \mathsf{getSignal}\left(\mathsf{getStrength}(pw_u) \right)$.

\ignore{\begin{algorithm}[h]
\caption{Delayed Signaling in Authentication}
\label{algo:delay}
\begin{algorithmic}[1]
\REQUIRE{$u, pw^{\prime}$}
\ENSURE{SUCCESS/FAIL}
\STATE $(u, salt_u, sig_u, h_u) \gets \mathsf{findRecord}(u)$ 
\STATE $h^{\prime} \gets H(salt_u, pw^{\prime})$
\IF{$h^{\prime} \neq h_u$} 
\RETURN FAIL
\ENDIF
\IF{$sig_u = \bot$ (no prior signal)}
\STATE $str_u \gets \mathsf{getStrength}(pw^{\prime})$
\STATE $y = \overset{\$}{\leftarrow} \mathsf{getSignal}(str_u)$
\STATE $\mathsf{updateRecord}$ as $(u,salt_u, sig_u=y, h_u)$
\ENDIF
\RETURN SUCCESS
\end{algorithmic}
\end{algorithm}
}

%% file: sections/adversary.tex
We adapt the economic model of \cite{BlockiD16} to capture the behavior of a rational attacker. We also make several assumptions: (1) there is a value $v_u$ for each password $pw_u$ that the attacker cracks; (2) the attacker is untargeted and that the value $v_u=v$ for each user $u \in U$; (3) by Kerckhoffs's principle, the password distribution $\mathcal{P}$ and the signaling matrix are known to the attacker.

\paragraph{Value/Cost Estimates} One can derive a range of estimates for $v$ based on black market studies e.g., Symantec reported that passwords generally sell for \$4---\$30~\cite{passwordBlackMarket} and \cite{stockley_2016} reported that Yahoo! e-mail passwords sold for $\approx \$1$. Similarly, we assume that the attacker pays a cost $k$ each time he evaluates the hash function $H$ to check a password guess. We remark that one can estimate $k\approx \$1 \times 10^{-7}$ if we use a memory-hard function \footnote{The energy cost of transferring 1GB of memory between RAM and cache is approximately $0.3J$ on an \cite{TCC:RenDev17}, which translates to an energy cost of $\approx\$3\times 10^{-8}$ per evaluation. Similarly, if we assume that our MHF can be evaluated in 1 second~\cite{Argon2,C:BHKLXZ19} then evaluating the hash function $6.3 \times 10^7$ times will tie up a $1$GB RAM chip for $2$ years. If it costs $\$5$ to rent a $1$GB RAM chip for $2$ years (equivalently purchase the RAM chip which lasts for $2$ years for $\$5$) then the capital cost is $\approx\$8\times 10^{-8}$. Thus, our total cost would be around $\$10^{-7}$ per password guess.}.   

\subsection{Adversary Utility: No Signaling}\label{sec:advnosignal}

We first discuss how a rational adversary would behave when is no signal is available (traditional hashing). We defer the discussion of how the adversary would update his strategy after observing a signal $y$ to the next section. In the no-signaling case, the attacker's strategy $(\pi,B)$ is given by an ordering $\pi$ over passwords $\mathbb{P}$ and a threshold $B$. Intuitively, this means that the attacker will check the first $B$ guesses in $\pi$ and then give up. The expected reward for the attacker is given by the simple formula $v \times \lambda(\pi, B)$, i.e., the probability that the password is cracked times the value $v$. Similarly, the expected guessing cost of the attacker is 
\begin{equation}\label{eq:cost1} 
C(k,\pi, B)= k\sum^B_{i=1} (1-\lambda(\pi, i-1)),
\end{equation}
Intuitively, $(1-\lambda(\pi, i-1))$ denotes the probability that the adversary actually has to check the $i$th password guess at cost $k$. With probability $\lambda(\pi,i-1)$ the attacker will find the password in the first $i-1$ guesses and will not have to check the $i$th password guess $pw_i^{\pi}$. Specially, we define $\lambda(\pi,0)=0$. The adversary’s expected utility is the difference of expected gain and expected cost, namely,
\begin{equation}\label{eq:utility}
U_{adv}\left(v,k,\pi, B \right)=v\cdot \lambda(\pi,B)-C(k, \pi, B).
\end{equation}
Sometimes we omit parameters in the parenthesis and just write $U_{adv}$ for short when the $v,k$ and $B$ are clear from context. 

\subsection{Optimal Attacker Strategy:  No Signaling} 
A rational adversary would choose $(\pi^*, B^*) \in \arg\max U_{adv}\left(v,k,\pi,B \right)$. It is easy to verify that the optimal ordering $\pi^*$ is always to check passwords in descending order of probability. The probability that a random user’s account is cracked is 
\begin{equation}
P_{adv}= \lambda(\pi^*, B^*).
\end{equation}
We remark that in practice $\arg\max U_{adv}\left(v,k,\pi,B \right)$ usually returns a singleton set $(\pi^*,B^*)$. If instead the set contains multiple strategies then we break ties adversarially i.e., $$P_{adv}= \max_{(\pi^*,B^*) \in  \arg\max U_{adv}\left(v,k,\pi,B \right)} \lambda(\pi^*, B^*).$$

%% file: sections/game.tex



We model the interaction between the authentication server (leader) and the adversary (follower) as a two-stage Stackelberg game. In a Stackelberg game, the leader moves first and then the follower may select its action after observing the action of the leader. 

In our setting the action of the defender is to commit to a signaling matrix $\mathbf{S}$ as well as the implementation of $\getstr$ which maps passwords to strength levels. The attacker responds by selecting a cracking strategy $( \vec{\pi}, \vec{B}) = \{ (\pi_0,B_0),\ldots,(\pi_{b-1},B_{b-1}) \}$. Intuitively, this strategy means that whenever the attacker observes a signal $y$ he will check the top $B_y$ guesses according to the ordering $\pi_y$.


\subsection{Attacker Utility}
If the attacker checks the top $B_y$ guesses according to the order $\pi_y$ then the attacker will crack the password with probability $\lambda(\pi_y,B_y;y)$. Recall that $\lambda(\pi_y,B_y;y)$ denotes the probability of the first $B_y$ passwords in $\pi_y$ according to the posterior distribution $\mathcal{P}_y$ obtained by applying Bayes Law after observing a signal $y$. Extrapolating from no signal case, the expected utility of adversary conditioned on observing the signal $y$ is 
\begin{equation}
  \label{eq:uadvsig}
  \begin{aligned}
&U_{adv}(v, k, \pi_y,B_y;\mathbf{S}, y)\\
&=v\cdot \lambda(\pi
_y, B_y;y)-\sum^{B_y}_{i=1} k\cdot \left(1-\lambda(\pi
_y, i-1;y)\right),
  \end{aligned}
\end{equation}
where $B_y$ and $\pi_y$ are now both functions of the signal $y$. Intuitively, $\left(1-\lambda(\pi_y, i-1;y)\right)$ denotes the probability that the attacker has to pay cost $k$ to make the $i$th guess.  We use $U_{adv}^{s}\left(v, k, \{\mathbf{S}, (\vec{\pi}, \vec{B})\}\right)$ to denote the expected utility of the adversary with information signaling, 

\begin{equation}
  \begin{aligned}
&U_{adv}^{s} \left(v, k, \{\mathbf{S}, (\vec{\pi}, \vec{B})\}\right) \\
& = \sum_{y \in [b]}\Pr[Sig = y]U_{adv}(v, k, \pi_y,B_y;\mathbf{S}, y) \ , 
  \end{aligned}
\end{equation}
where \[ Pr[Sig = y] = \sum_{i \in [b]} \Pr_{pw \sim \mathcal{P}} [\mathsf{getStrength}(pw)=i] \cdot S[i,y] \ . \]

\subsection{Optimal Attacker Strategy}
Now we discuss how to find the optimal strategy $(\vec{\pi}^*,\vec{B}^*)$. Since the attacker's strategies in reponse to different signals are independent. It suffices to find $(\pi_y^*, B_y^*) \in \arg\max_{B_y,\pi_y}  U_{adv}(v, k, \pi_y,B_y;y)$ for each signal $y$. We first remark that the adversary can obtain the optimal checking sequence $\pi_y^*$ for $pw_u$ associated with signal $y$ by sorting all $pw\in\mathcal{P}$ in descending order of posterior probability according to the posterior distribution $\mathcal{P}_y$. 

Given the optimal guessing order $\pi_y^*$, the adversary can determine the optimal budget $B_{y}^*$ for signal $y$ such that $B_{y}^* = \arg\max_{B_y}  U_{adv}(v, k, \pi_y^*,B_y;y)$. Each of the password distributions we analyze has a compact representation allowing us to apply techniques from \cite{DAHash} to further speed up the computation of the attacker's optimal strategy $\pi_y^*$ and $B_y^*$ --- see discussion in the appendix.  

We observe that an adversary who sets $\pi_y = \pi$ and $B_y = B$  for all $y\in [b]$ is effectively ignoring the signal and is equivalent to an adversary in the no signal case. Thus,
\begin{equation}
\max_{\vec{\pi}, \vec{B}}U_{adv}^{s} \left(v, k, \{\mathbf{S}, (\vec{\pi}, \vec{B})\}\right) \geq\max_{\pi, B} U_{adv} (v, k, \pi, B), \;\forall \mathbf{S}, 
\end{equation}

implying that adversary's expected utility will never decrease by adapting its strategy according to the signal.

\subsection{Optimal Signaling Strategy}
Once the function $\mathsf{getStrength}()$ is fixed we want to find the optimal signaling matrix $\mathbf{S}$. We begin by introducing the defender's utility function. Intuitively, the defender wants to minimize the total number of cracked passwords. 

Let  $P_{adv}^{s}\left(v, k, \mathbf{S}\right)$ denote the expected adversary success rate with information signaling when playing with his/her optimal strategy, then 

\begin{equation}
P_{adv}^{s}\left(v, k, \mathbf{S}\right) = \sum_{y \in SL}\Pr[Sig = y] \lambda(\pi_y^*, B_y^*;\mathbf{S}, y),
\end{equation}
where $(\pi_y^*, B_y^*)$ is the optimal strategy of the adversary when receiving signal $y$, namely, $$(\pi_y^*, B_y^*) = \arg \max_{\pi_y, B_y} U_{adv}(v, k, \pi_y,B_y;\mathbf{S}, y).$$ If $\arg \max_{\pi_y, B_y} U_{adv}(v, k, \pi_y,B_y;y)$ returns a set, we break ties adversarially.

The objective of the server is to minimize $P_{adv}^{s}\left(v, k, \mathbf{S}\right)$, therefore we define
\begin{equation}
  U_{ser}^{s}\left(v, k, \{\mathbf{S}, (\vec{\pi}^*, \vec{B}^*)\}\right) = -P_{adv}^{s}\left(v, k, \mathbf{S}\right).
\end{equation}

Our focus of this paper is to find the optimal signaling strategy, namely, the signaling matrix $\mathbf{S}^*$ such that $\mathbf{S}^* = \arg\min _{\mathbf{S}} P_{adv}^{s}\left(v, k, \mathbf{S}\right)$.
Finding the optimal signaling matrix $\mathbf{S}^*$ is equivalent to solving the mixed strategy Subgame Perfect  Equilibrium (SPE) of the Stackelberg game.
At SPE no player has the incentive to derivate from his/her strategy. Namely,
\vspace{-0.15cm}
\begin{equation}\small
\begin{cases}
  U_{ser}^{s}\left(v, k, \{\mathbf{S}^*, (\vec{\pi}^*, \vec{B}^*)\}\right) \geq  U_{ser}^{s}\left(v, k, \{\mathbf{S}, (\vec{\pi}^*, \vec{B}^*)\}\right),\forall \mathbf{S},\\
U_{adv}^{s} \left(v, k, \{\mathbf{S}^*, (\vec{\pi}^*, \vec{B}^*)\}\right) \geq U_{adv}^{s} \left(v, k, \{\mathbf{S}^*, (\vec{\pi}, \vec{B})\}\right), \forall(\vec{\pi},\vec{B}).
\end{cases}
\end{equation}

Notice that a signaling matrix of dimension $a\times b$ can be fully specified by $a(b-1)$ variables since the elements in each row sum up to 1. Fixing $v$ and $k$, we define $f:\mathbb{R}^{a(b-1)}\rightarrow 
\mathbb{R}$ to be the map from $\mathbf{S}$ to $P_{adv}^{s}\left(v, k, \mathbf{S}\right)$.  Then we can formulate the optimization problem as
\vspace{-0.15cm}
\begin{mini}|s| 
{\mathbf S}{ f (s_{0,0},\ldots s_{0,(b-2)}, \ldots, s_{(a-1),0},s_{(a-1),(b-2)})}
{}{}
\addConstraint{0 \leq s_{i,j} \leq 1, \; \forall 0 \leq i\leq a-1,\; 0\leq j\leq b-2}
\addConstraint{\sum_{j=0}^{b-2} s_{i,j} \leq 1,\; \forall 0\leq i\leq a-1}.
\label{optProblem}
\end{mini}
\vspace{-0.15cm}
The feasible region is a $a(b-1)$-dimensional probability simplex. Notice that in 2-D ($a = b = 2$), the second constraint would be equivalent to the first constraint. In our experiments we will treat $f$ as a black box and use derivative-free optimization methods to find good signaling matrices $\sm$.

%% file: sections/theoretical_example.tex
Having presented our Stackelberg Game model for information signaling we now give an (admittedly contrived) example of a password distribution where information signaling can dramatically reduce the percentage of cracked passwords. We assume that the attacker has value $v = 2k + \epsilon$ for each cracked password where the cost of each password guess is $k$ and $\epsilon>0$ is a small constant.

\paragraph{Password Distribution} Suppose that $\mathbb{P}=\{\text{pw}_i\}_{i \geq 1}$ and that each password $pw_i$ has probability $2^{-i}$ i.e., $\Pr\limits_{pw \sim \mathcal{P}}\left[pw= i\right] = 2^{-i}$. The weakest password $pw_1$ would be selected with probability $1/2$. 

\paragraph{Optimal Attacker Strategy without Signaling}
By checking passwords in descending order of probability (the checking sequence is $\pi$) the adversary has an expected cost
\vspace{-0.1cm}
\begin{align*}
C(k,\pi,B) &= k\sum_{i=1}^{B} i \times 2^{-i} + 2^{-B} \times B \times k = 2k(1-2^{-B}).
\end{align*}
The expression above is equivalent to equation \eqref{eq:cost1}, the attacker succeeds in $i$th guess with probability $2^{-i}$ at the cost of $ik$; the attacker fails after making $B$ guess with probability $2^{-B}$ at the cost of $Bk$. The the expected gain is
\vspace{-0.1cm}
$$
R(v,k,\pi,B) = v \sum_{i=1}^{B} i 2^{-i} = v\left(1-2^{-B}\right).
$$
\vspace{-0.1cm}
Therefore the expected utility is
\begin{small}
\vspace{-0.1cm}
$$U_{adv}(v,k,\pi,B) = R(v,k,B) - C(k,\pi,B) = (v-2k)\left(1- 2^{-B}\right).$$
\end{small}
A profit-motivated adversary is interested in calculating $B^* = \underset{B}{\text{argmax }} U_{adv}(v,k,\pi,B)$. With our sample distribution we have
\[
B^* = 
\begin{cases}
0, & \mbox{if } v < 2k,\\
\infty & \mbox{if } v \geq 2k.
\end{cases}
\]

Since we assume that $v = 2k + \epsilon > 2k$ the attackers optimal strategy is $B^* = \infty$ meaning that $100\%$ of passwords will be cracked.

\paragraph{Signaling Strategy}
Suppose that $\mathsf{getStrength}$ is define such that $\mathsf{getStrength}(pw_1)=0$ and $\mathsf{getStrength}(pw_i)=1$ for each $i > 1$. Intuitively, the strength level is $0$ if and only if we sampled the weakest password from the distribution. Now suppose that we select our signaling matrix  $$\mathbf{S} = \begin{bmatrix}
1/2 & 1/2 \\
0 & 1 
\end{bmatrix}\ ,$$
such that $\Pr[Sig = 0 ~|~ pw = pw_1] = \frac{1}{2} = \Pr[Sig = 1 ~|~ pw = pw_1]$ and $\Pr[Sig = 1 ~|~ pw \neq pw_1] = 1$. 

\paragraph{Optimal Attacker Strategy with Information Signaling}
We now analyze the behavior of a rational attacker under signaling when given this signal matrix and password distribution assuming that the attacker's value $v = 2k + \epsilon$ is the same. 

We first note that attacker observes the signal $Sig = 0$ we know for sure that the user selected the most common password as $\Pr[pw = pw_1 ~|~ Sig = 0] = 1$ so as long as $v \geq k$ the attacker will crack the password.   

Next we consider the case that the attacker observes the signal $Sig = 1$. We have the following posterior probabilities for each of the passwords in the distribution:

\begin{small}
\begin{align*}
\Pr[pw = pw_1 ~|~ Sig = 1] 
&= \frac{0.5 \times 0.5}{0.75} = \frac{1}{3},\\
\Pr[pw = pw_i, i > 1 ~|~ Sig = 1] 
&= \frac{1 \times 2^{-i}}{0.75} = \frac{4 \times 2^{-i}}{3}.
\end{align*}
\end{small}
Now we compute the attacker's expected costs conditioned on $Sig = 1$. 
\begin{equation*}
\begin{aligned}
C(k,\pi,B;\mathbf{S},1) &= k\left(\frac{1}{3} + \frac{4}{3}\sum_{i=2}^{B} i*2^{-i}\right)+kB\left(\frac{2}{3}-\frac{4}{3}\left(\sum_{i=2}^{B}2^{-i}\right)\right) \\
&=k\left(\frac{7}{3} - \frac{2^{3-B}}{3}\right),
\end{aligned}
\end{equation*}
The expected gain of the attacker is
\begin{equation*}
\begin{aligned}
R(v,k,\pi,B;\mathbf{S},1) = v\left(\frac{1}{3} + \frac{4}{3}\sum_{i=2}^B 2^{-i}\right) = v\left(1 - \frac{2^{2-B}}{3}\right),
\end{aligned}
\end{equation*}
Thus, the attacker's utility is given by:
\begin{equation*}
\begin{aligned}
U_{adv}(v,k,\pi,B;\mathbf{S},1) &= R(v,k,\pi,B;\mathbf{S},1) - C(k,\pi,B;\mathbf{S},1) \\
&= v\left(1 - \frac{2^{2-B}}{3}\right) - k\left(\frac{7}{3} - \frac{2^{3-B}}{3}\right).
\end{aligned}
\end{equation*}

Assuming that $\epsilon < \frac{1}{3}k$ the attacker will have negative utility $U_{adv}(2k + \epsilon,k,\pi,B;\mathbf{S},1) < 0$ whenever $B>1$. Thus, when the signal is $Sig=1$ the optimal attacker strategy is to select $B^* = 0$ (i.e., don't attack) to ensure zero utility. In particular, the attacker cracks the password if and only if $Sig=0$ which happens with probability $1-\Pr[Sig=1]=0.25$ since $\Pr[Sig=1] =$ $\Pr[pw=pw_1]\Pr[Sig=1 ~|~ pw = pw_1] + \Pr[pw \neq pw_1]\Pr[Sig=1 ~|~ pw \neq pw_1] $ $= \frac{3}{4} $. Thus, the attacker will only crack $25\%$ of passwords when $v=2k+\epsilon$\footnote{If the attacker's value was increased to $v = 4k$ then the attacker would crack $100\%$ of passwords since we would have $U_{adv}(4k,k,\pi,B;\mathbf{S},1)= k\left(\frac{5}{3} - \frac{2^{3-B}}{3}\right)$ which is maximized at $B^* = \infty$. In this case the defender would want to look for a different password signaling matrix. }.

\paragraph{Discussion} In our example an attacker with value $v = 2k + \epsilon$ cracks $100\%$ of passwords when we don't use information signaling. However, if our information signaling mechanism (above) were deployed, the attacker will only crack $25\%$ of passwords --- a reduction of $75\%$! Given this (contrived) example it is natural to ask whether or not information signaling produces similar results for more realistic password distributions. We explore this question in the next sections.

%% file: sections/implementation.tex

We now describe our empirical experiments to evaluate the performance of information signaling. Fixing the parameters $v,k,a,b$, a password distribution $\mathcal{D}$ and the strength oracle  $\mathsf{getStrength}(\cdot)$ we define a procedure $\osm \leftarrow\mathsf{genSigMat}(v,k,a, b, \mathcal{D})$ which uses derivate-free optimization to solve the optimization problem defined in equation \eqref{optProblem} and  find a good generate a signaling matrix  $\osm$ of dimension $a \times b$. Similarly, given a signaling matrix $\osm$ we define a procedure $\mathsf{evaluate}(v,k,a,b,\osm, \mathcal{D})$ which returns the percentage of passwords that a rational adversary will crack given that the value of a cracked password is $v$, the cost of checking each password is $k$. To simulate settings where the defender has imperfect knowledge of the password distribution we use different distributions $\mathcal{D}_1$ (training) and $\mathcal{D}_2$ (evaluation) to generate the signaling matrix $\osm \leftarrow\mathsf{genSigMat}(v,k,a, b, \mathcal{D}_1)$ and evaluate the success rate  of a rational attacker $\mathsf{evaluate}(v,k,a,b,\osm, \mathcal{D}_2)$. We can also set $\mathcal{D}_1=\mathcal{D}_2$ to evaluate our mechanism under the idealized setting in which defender has perfect knowledge of the distribution.

In the remainder of this section we describe how the oracle $\getstr$ is implemented in different experiments, the password distribution(s) derived from empirical password datasets and how we implement $\mathsf{genSigMat}()$.

\subsection{Password Distribution} 

We evaluate the performance of our information signaling mechanism using 9 password datasets: Bfield (0.54 million), Brazzers ($N=0.93$ million), Clixsense (2.2 million), CSDN (6.4 million), LinkedIn (174 million), Neopets (68.3 million), RockYou (32.6 million), 000webhost (153 million) and Yahoo! (69.3 million). The Yahoo! frequency corpus ($N \approx 7 \times 10^7$) was collected and released with permission from Yahoo! using differential privacy~\cite{NDSS:BloDatBon16} and other privacy-preserving measures \cite{SP:Bonneau12}. All the other datasets come from server breaches. 

\paragraph{Empirical Distribution}  For all 9 datasets we can derive an empirical password distribution $\mathcal{D}_e$ where $\Pr_{pw\sim \mathcal{D}_e}[pw_i]= f_i/N$. Here, $N$ is the number of users in the dataset and $f_i$ is the number of occurences of $pw_i$ in the dataset. We remark that for datasets like Yahoo! and LinkedIn where the datasets only include frequencies $f_i$ without the original plaintext password we can derive a distribution simply by generating unique strings for each password. The empirical distribution is useful to analyze the performance of information signaling when the password value $v$ is small this analysis will be  less accurate for larger values of $v$ i.e., once the rational attacker has incentive to start cracking passwords with lower frequency. Following an approach taken in \cite{DAHash}, we use Good-Turing frequency estimation
to identify and highlight regions of uncertainty where the CDF for the empirical distribution might significantly diverge from the real password distribution. To simulate an attacker with imperfect knowledge of the distribution we train a differentially private Count-Mean-Min-Sketch. In turn, the Count-Sketch is used to  derive a distribution $\mathcal{D}_{train}$, to  implement $\getstr$ and to generate the signaling matrix  $\osm \leftarrow\mathsf{genSigMat}(v,k,a, b, \mathcal{D}_{train})$ (see details below).

\paragraph{Monte Carlo Distribution}
To derive the Monte Carlo password distribution from a dataset we follow a process from \cite{DAHash}. In particular, we subsample passwords $D_s \subseteq D$ from the dataset and derive guessing numbers $\#guessing_m(pw)$ for each $pw \in D_s$. Here, $\#guessing_m(pw)$ denotes the number of guesses needed to crack $pw$ with a state of the art password cracking model $m$ e.g.,  Probabilistic Context-Free Grammars~\cite{SP:WAMG09,SP:KKMSVB12,NDSS:VerColTho14}, Markov models~\cite{NDSS:CasDurPer12,Castelluccia2013,SP:MYLL14,USENIX:USBCCKKMMS15}, and neural networks~\cite{USENIX:MUSKBCC16}. We used the password guessing service~\cite{USENIX:USBCCKKMMS15} to generate the guessing numbers for each dataset. We then fit our distribution to the guessing curve i.e., fixing thresholds $t_0=0 < t_1 < t_2 \ldots$ we assign any password $pw$ with $t_{i-1}< \min_m\{\#guessing_m(pw)\}\leq t_i$  to have probability $\frac{g_i}{|D_s|(t_i-t_{i-1})}$ where $g_i$ counts the number of sampled passwords in $D_s$ with guessing number between $t_{i-1}$ and $t_i$. Intuitively, the Monte Carlo distribution $\mathcal{D}_m$  models password distribution from the attacker's perspective. One drawback is that the distribution would change if the attacker were to develop an improved password cracking model.

 We extract Monte Carlo distribution from 6 datasets (Bfield, Brazzers, Clixsense, CSDN, Neopets, 000webhost) for which we have plain text passwords so that we can query Password Guessing Service \cite{USENIX:USBCCKKMMS15} about password guessing numbers. In the imperfect knowledge setting we repeated the process above twice for each dataset with disjoint sub-samples to derive two distributions $\mathcal{D}_{train}$ and $D_{eval}$.

\subsection{Differentially Private Count-Sketch}
When using the empirical distribution $\mathcal{D}_{e}$ for evaluation we evaluate the performance of an imperfect knowledge defender who trains a differentially private Count-Mean-Min-Sketch. As users register their accounts, the server can feed passwords into a Count-Mean-Min-Sketch initialized with Laplace noise to ensure differential privacy. 


 When working with empirical distributions in an imperfect knowledge setting we split the original dataset $D$ in half to obtain $D_{1}$ and $D_{2}$. Our noise-initialized Count-Mean-Min-Sketch is trained with $D_{1}$. We fix the width $d_w$ (resp. depth $d_s$) of our count sketch to be $d_w=10^8$ (resp. $d_s=10$) and add Laplace Noise with scaling factor $b= d_s/{\epsilon_{pri}}=5$ to preserve $\epsilon_{pri}=2$-differential privacy. Since we were not optimizing for space we set the number of columns $d_w$ to be large to minimize the probability of hash collisions and increase the accuracy of frequency estimation. Each cell is encoded by an 4-byte \textrm{int} type so the total size of the sketch is 4 GB. 
 
We then use this count sketch along with $D_2$ to extract a noisy distribution $\mathcal{D}_{train}$. In particular, for every $pw\in D_{2}$ we query the the count sketch to get $\tilde{f}_{pw}$, a noisy estimate of the frequency of $pw$ in $D_2$ and set $\Pr_{\mathcal{D}_{train}}[pw] \doteq \frac{\tilde{f}_{pw}}{\sum_{w \in D_2} \tilde{f}_{w}}$. We also use the Count-Mean-Min Sketch as a frequency oracle in our implementation of $\getstr$ (see details below). We then use $\mathcal{D}_{train}$ to derive frequency thresholds for $\getstr$ and to generate the signaling matrix $\osm = \mathsf{genSigMat}(v,k,a,b, \mathcal{D}_{train})$. Finally we evaluate results on the original empirical distribution $\mathcal{D}_e$ for the original dataset $D$  i.e., $P_{adv}^s = \mathsf{evaluate}(v,k,a,b, \osm, \mathcal{D}_e)$. 
 \vspace{-0.2cm}
 
\subsection{Implementing $\getstr$} 

Given a distribution $\mathcal{D}$ and a frequency oracle $\mathcal{O}$ which outputs $f(pw)$ in the perfect knowledge setting and an estimate of frequency $\hat{f}(pw)$ in the imperfect knowledge setting, we can specify $\getstr$ by selecting thresholds  $x_1 > \ldots > x_{a-1} > x_a = 1$. In particular, if $x_{i+1} \leq \mathcal{O}(pw) < x_i$ then $\getstr{pw}=i$ and if $\mathcal{O}(pw) \geq x_1$ then $\getstr{pw}=0$. Let $Y_i\doteq \Pr_{pw \sim \mathcal{D}}[x_{i} \leq \getstr{pw} < x_{i-1}]$ for $i > 1$ and $Y_1 = \Pr_{pw \sim \mathcal{D}}[\getstr{pw} > x_1]$. We fix the thresholds $x_1 \geq \ldots \geq x_{a-1}$ to (approximately) balance the probability mass of each strength level i.e., to ensure that $Y_{i} \approx Y_{j}$. 
In imperfect (resp. perfect) knowledge settings we use $\mathcal{D}=\mathcal{D}_{train}$ (resp. $\mathcal{D}=\mathcal{D}_{eval}$) to select the thresholds.

\subsection{Derivative-Free Optimization} Given a value $v$ and hash cost $k$ we want to find a signaling matrix which optimizes the defenders utility. Recall that this is equivalent to minimizing the function $f(\sm) = \mathsf{evaluate}(v,k,a,b,\sm, \mathcal{D}) $ subject to the constraints that $\sm$ is a valid signaling matrix. In our experiments we will treat $f$ as a black box and use derivative-free optimization methods to find good signaling matrices $\osm$. 

 In our experiment, we choose BITmask Evolution OPTimization (BITEOPT) algorithm \cite{biteopt2021} to compute the quasi-optimal signaling matrix $\osm$. BITEOPT is a free open-source stochastic non-linear bound-constrained derivative-free optimization method (heuristic or strategy). BiteOpt took 2nd place (1st by sum of ranks) in BBComp2018-1OBJ-expensive competition track \cite{BBComp}.

In each experiment we  use BITEOPT with $10^4$ iterations to generate signaling matrix $\osm$ for each different  $v/C_{max}$ ratio, where $C_{max}$ is server's maximum authentication cost satisfying $k \leq C_{max}$. We refer to the procedure as  $\osm \leftarrow\mathsf{genSigMat}(v,k,a, b, \mathcal{D}_1)$ . 


%% file: sections/experiments.tex

We describe the results of our experiments. In the first batch of experiments we evaluate the performance of information signaling against an offline and an online attacker where the ratio $v/C_{max}$ is typically much smaller.  
\vspace{-0.1cm}
\subsection{Password Signaling against Offine Attacks}

 We consider four scenarios using the empirical/Monte Carlo distribution in a setting where the defender has perfect/imperfect knowledge of the distribution.

\subsubsection{Empirical Distribution}
From each password dataset we derived an empirical distribution $\mathcal{D}_{e}$ and set $\mathcal{D}_{eval} = \mathcal{D}_{e}$. In the perfect knowledge setting we also set $\mathcal{D}_{train}=\mathcal{D}_e$ while in the imperfect knowledge setting we used a Count-Min-Mean Sketch to derive $\mathcal{D}_{train}$ (see details in the previous section). 

We fix dimension of signaling matrix to be 11 by 3 (the server issues 3 signals for 11 password strength levels) and compute attacker's success rate for different value-to-cost ratios $v/C_{max} \in \{i \times 10^j:1\leq i\leq 9,\, 3\leq j \leq 7\} \cup \{(i+0.5) \times 10^j:1\leq i\leq 9,\, 6\leq j \leq 7\}$ . In particular, for each value-to-cost ratio $v/C_{max}$ we run $\osm \leftarrow \mathsf{genSigMat}(v,k,a, b, \mathcal{D}_e)$ to generate a signaling matrix and then run $ \mathsf{evaluate}(v,k,a, b, \osm, \mathcal{D}_e)$ to get the attacker's success rate. The same experiment is repeated for all 9 password datasets. We plot the attacker's success rate vs. $v/C_{max}$ in Fig. \ref{fig:empirical}. Due to space limitations Fig. \ref{fig:empirical} only shows results for 6 datasets --- additional plots can be found in Fig \ref{fig:empiricalExtra} in the Appendix.  

We follow the approach of \cite{DAHash}, highlighting the uncertain regions of the plot where the cumulative density function of the empirical distribution might diverge from the real distribution. In particular, the red (resp. yellow) region indicates $E > 0.1$ (resp. $E > 0.01$) where $E$ can be interpreted as an upper bound on the difference between the two CDFs.

\input{figs/empirical.tex}

Fig. \ref{fig:empirical} demonstrates that information signaling reduces the fraction of cracked passwords. The mechanism performs best when the defender has perfect knowledge of the distribution (blue curve), but even with imperfect knowledge there is still a large advantage. For example, for the neopets dataset when $v/C_{max}= 5\times 10^6$ the percentage of cracked passwords is reduced from $44.6\%$ to $36.9\%$ (resp. $39.1\% $) when the defender has perfect (resp. imperfect) knowledge of the password distribution. Similar results hold for other datasets.   
The green curve (signaling with imperfect knowledge) curve generally lies in between the black curve (no signaling) and the blue curve (signaling with perfect knowledge), but sometimes has an adverse affect affect when $v/C_{max}$ is large. This is because the noisy distribution will be less accurate for stronger passwords that were sampled only once.

\paragraph{Which accounts are cracked?}  As Fig \ref{fig:empirical} demonstrates information signaling can substantially reduce the overall fraction of cracked passwords i.e., many previously cracked passwords are now protected. It is natural to ask whether there are any unlucky users $u$ whose password is cracked after information signaling {\em even though} their account was safe before signaling. Let $X_u$ (resp. $L_u$) denote the event that user $u$ is unlucky (resp. lucky) i.e., a rational attacker would originally not crack $pw_u$, but after information signaling the account is cracked. We measure $E[X_u]$ and $E[L_u]$ (See Fig. \ref{fig:unlucky}) for various $v/C_{max}$  values under each dataset. Generally, we find that the fraction of unlucky users $E[X_u]$  is  small in most cases e.g. $\leq 0.04$.  For example, when $v/k=2\times 10^7$ we have that $E[X_u] \approx 0.03\%$ and   $E[L_u] \approx 6\%$ for LinkedIn. In all instances the net advantage $E[L_u] - E[X_u]$ remains positive. 
\input{figs/unlucky.tex}
We remark that the reduction in cracked passwords does not come from persuading the attacker to crack weak passwords though the attacker might shift his attention.
The shift of attacker’s attention is directionless, not necessarily towards weaker passwords.  The contribution of attention shift to reduction in cracked passwords is very small since the passwords ordering of posterior distribution upon receiving a signal is very close to that of prior distribution, which means the attacker cracks passwords (almost) in the same order whether given the signal or not. Strength signaling works mainly because the attacker would like to save cost by making less futile guesses.

\paragraph{Robustness} We also evaluated the robustness of the signaling matrix when the defender's estimate of the ratio $v/C_{max}$  is inaccurate. In particular, for each dataset we generated the signaling matrix $\sm(10^5)$ (resp. $\sm(10^6)$)  which was optimized with respect to the ratio $v/C_{max} =10^5$ (resp. $v/C_{max} =10^6$) and evaluated the performance of both signaling matrices against an attacker with different $v/C_{max}$  ratios. We find that password signaling is tolerant even if our estimate of $v/k$ is off by a small multiplicative constant factor e.g., $2$. For example, in  Fig. \ref{fig:empiricalneopets} the signaling matrix $\mathbf{S}(10^6)$ outperforms the no-signaling case even when the real $v/C_{max}$  ratio is as large as $2 \times 10^6$. In the ``downhill'' direction,  even if the estimation of $v/k$ deviates from its true value up to $5\times 10^5$ at anchor point $10^6$  it is still advantageous for the server to deploy password signaling.

\subsubsection{Monte Carlo Distribution}
We use the Monte Carlo distribution to evaluate information signaling when $v/C_{max}$ is large. In particular, we subsample 25k passwords from each datast for which we have plain text passwords (excluding Yahoo! and LinkedIn) and obtain guessing numbers from the Password Guessing Service. Then we split our 25k subsamples in half to obtain two guessing curves and we extract two Monte Carlo distributions $\mathcal{D}_{train}$ and $\mathcal{D}_{eval}$ from these curves (see details in the last section). In the perfect knowledge setting the signaling matrix is both optimized and tested on $\mathcal{D}_{eval}$ i.e., $\osm = \mathsf{genSigMat}(v,k,a,b, \mathcal{D}_{eval})$, $P_{adv}^s = \mathsf{evaluate}(v,k,a,b, \osm, \mathcal{D}_{eval})$. In the imperfect knowledge setting the signaling matrix is tuned on $\mathcal{D}_{train}$ while the attacker's success rate is evaluated on $\mathcal{D}_{eval}$.  
 One advantage of simulating Monte Carlo distribution is that it allows us to evaluate the performance of information signaling against state of the art password cracking models when the $v/C_{max}$ is large. We consider $v/C_{max} \in \{i*10^j:1\leq i\leq 9,\, 5\leq j \leq 10\}$ in performance evaluation for Monte Carlo distribution. As before we set $a=11$ and $b=3$ so that the signaling matrix is in dimension of $11 \times 3$. We present our results in Fig. \ref{fig:monte}. 

Fig. \ref{fig:monte} shows that information signaling can significantly reduce the number of cracked passwords. In particular, for the neopets dataset when $v/C_{max}=6 \times 10^7$ the number of cracked passwords is reduced from $52.2\% $ to $40\%$ (resp. $43.8\%$) when the defender has perfect (resp. imperfect) knowledge of the distribution. The green curve (signaling with imperfect knowledge) generally lies between the black curve (no signaling) and the blue curve (signaling with perfect information) though we occasionally find points where the green curve lies slightly above the black curve.

\input{figs/monte.tex}

\vspace{-0.2cm}
\subsection{Password Signaling against Online Attacks}
We can extend the experiment from password signaling with perfect knowledge to an online attack scenario. One common way to throttle online attackers is to require the attacker to solve a CAPTCHA challenge \cite{EC:vBHL03}, or provide some other proof of work (PoW), after each incorrect login attempt~\cite{CCS:PinSan02}. One advantage of this approach is that a malicious attacker cannot lockout an honest user by repeatedly submitting incorrect passwords \cite{wolverton2002hackers}. However, the solution also allows an attacker to continue trying to crack the password as long as s/he is willing to continue paying the cost to solve the CAPTCHA/PoW challenges. Thus, information signaling could be a useful tool to mitigate the risk of online attacks. 

When modeling a rational online password we will assume that $v/C_{max} \leq 10^5$ since the cost to pay a human to solve a CAPTCHA challenge (e.g., $\$10^{-3}$ to $10^2$ \cite{USENIX:MLKMVS10}) is typically much larger than the cost to evaluate a memory-hard cryptographic hash function (e.g., $\$10^{-7}$). Since $v/C_{max} \leq 10^5$ we use the empirical distribution to evaluate the performance of information signaling against an online attacker. In the previous subsection we found that the uncertain regions of the curve started when $v/C_{max} \gg 10^5$ so the empirical distribution is guaranteed to closely match the real one. 

Since an online attacker will be primarily focused on the most common passwords (e.g., top $10^3$ to $10^4$) we modify $\getstr$ accordingly. We consider two modifications of $\getstr$ which split passwords in the top $10^3$ (resp. $10^4$) passwords into $11$ strength levels. By contrast, our prior implementation of $\getstr$ would have placed most of the top $10^3$ passwords in the bottom two strength levels. As before we fix the signaling matrix dimension to be $11\times 3$. Our results are shown in Fig. \ref{fig:online}. Due to space limitations the results for 6 datasets are in Fig. \ref{fig:onlineExtra} in the appendix.

Our results demonstrate that information signaling can be an effective defense against online attackers as well. For example, in Fig. \ref{fig:onlinebrazzers}, when $v/C_{max} =9\times 10^4$, our mechanism reduces the fraction of cracked passwords from $20.4\%$ to just $15.3\%$. Similar, observations hold true for other datasets.

We observe that the red curve (partitioning the top $10^3$ passwords into $11$ strength levels) performs better than the blue curve (partitioning the top $10^3$ passwords into $11$ strength levels) when $v/k$ is small e.g.,  $v/C_{max} < 2\times10^4$ in Fig. \ref{fig:onlinebrazzers}). The blue curve performs better when $v/C_{max}$  is larger. Intuitively, this is because we want to have a fine-grained partition for the weaker (top $10^3$) passwords that the adversary might target when $v/C_{max}$  is small. 

\paragraph{Implementing Password Signaling} One naive way to implement password signaling in an online would simply be to explicitly send back the signal noisy signal $sig_u$ in response to any incorrect login attempt. As an alternative we propose a solution where users with a weaker signal $sig_u$ are throttled more aggressively. For example, if $sig_u$ indicates that the password is strong then it might be reasonable to allow for $10$ consecutive incorrect login attempts before throttling the account by requiring the user to solve a CAPTCHA challenge before every login attempt. On the other hand if the signal $sig_u$ indicates that the password is weak the server might begin throttling after just $3$ incorrect login attempts. The attacker can indirectly infer the signal $sig_u$ by measuring how many login attempts s/he gets before throttling begins. This solution might also provide motivation for users to pick stronger passwords. 

\input{figs/online.tex}
\vspace{-0.1cm}
\subsection{Discussion}
While our experimental results are positive, we stress that there are several questions that would need to be addressed before we recommend deploying information signaling to protect against offline attacks. 
\begin{itemize}
\item Can we accurately predict the value to cost ratio $v/C_{max}$? Our results suggest that information signaling is useful even when our estimates deviate by a factor of $2$. However, if our estimates are wildly off then information signaling could be harmful.
\item  While information signaling reduced the total number of cracked passwords a few unlucky users might be harmed i.e., instead of being deterred the unlucky signal helps the rational attacker to crack a password that they would not otherwise have cracked. The usage of password signaling raises important ethical and societal questions. How would users react to such a solution knowing that they could be one of the unlucky users? One possible way to address these concerns would be to allow user's to opt in/out of information signaling. However, each user $u$ would need to make this decision without observing their signal. Otherwise the decision to opt in/out might be strongly correlated with the signal allowing the attacker to perform another Bayesian update. Another possible way to address these concerns would be to modify the objective function (eq \ref{optProblem}) to penalize solutions with unlucky users.     
\item Can we analyze the behavior of rational targeted attackers? We only consider an untargeted attacker. In some settings, an attacker might place a higher value on some passwords e.g., celebrity accounts. Can we predict how a targeted attacker would behave if the value $v_{u}$ varied from user to user?  
Similarly, a targeted adversary could exploit demographic and/or biographical knowledge to improve password guessing attacks e.g., see~\cite{CCS:WZWYH16}. 

\end{itemize}

%% file: figs/empirical.tex
\tikzEmpirical
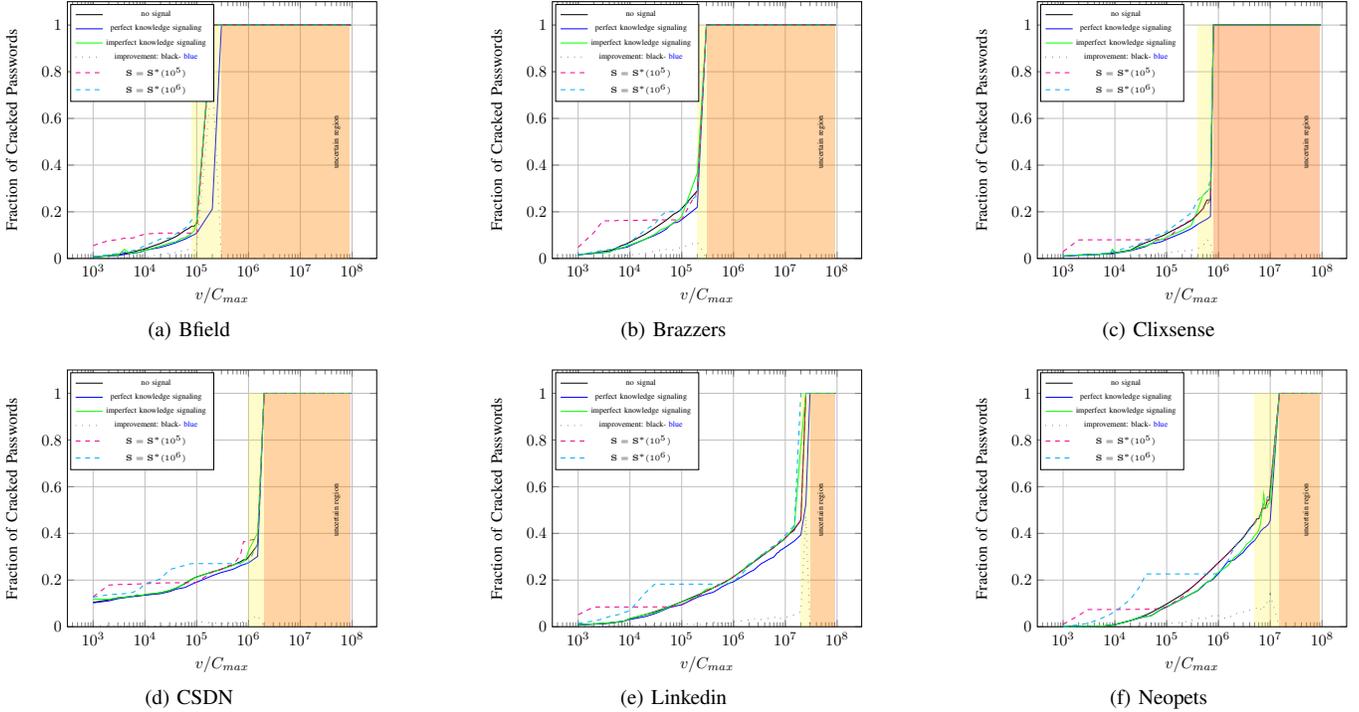
\begin{figure*}[ht]\centering
\subfloat[Bfield]{
\begin{tikzpicture}[scale=0.6]
\begin{semilogxaxis}[ymin=0]

\addplot+[stack plots=y] file {./figs/empirical/perfect/successrate/bfield.dat};
\addplot file {./figs/empirical/perfect/successrate/bfield11by3.dat};
\addplot file {./figs/empirical/imperfect/successrate/bfield11by3_test.dat};
\addplot+[stack plots=y, stack dir=minus] file {./figs/empirical/perfect/successrate/bfield11by3.dat};

\addplot file {./figs/empirical/perfect/successrate/bfield_1e5.dat};
\addplot file {./figs/empirical/perfect/successrate/bfield_1e6.dat};

\path[name path=B] (axis cs:(9 * 1e7,0) -- (axis cs:(9 * 1e7,1);
\path[name path=A] (axis cs:(300000,0) -- (axis cs:(300000,1);
\tikzfillbetween[of=A and B, on layer=main]{red, opacity=0.2};

\path[name path=A1] (axis cs:(80000,0) -- (axis cs:(80000,1);
\tikzfillbetween[of=A1 and B, on layer=main]{yellow, opacity=0.2};

\node at (axis cs:5*1e7,0.5) {\rotatebox{90}{\tiny uncertain region}};
\end{semilogxaxis}
\end{tikzpicture}
\label{fig:empiricalbfield}
}
\hfill
\subfloat[Brazzers]{
\begin{tikzpicture}[scale=0.6]
\begin{semilogxaxis}[ymin=0]

\addplot+[stack plots=y] file {./figs/empirical/perfect/successrate/brazzers.dat};
\addplot file {./figs/empirical/perfect/successrate/brazzers11by3.dat};
\addplot file {./figs/empirical/imperfect/successrate/brazzers11by3_test.dat};
\addplot+[stack plots=y, stack dir=minus] file {./figs/empirical/perfect/successrate/brazzers11by3.dat};

\addplot file {./figs/empirical/perfect/successrate/brazzers_1e5.dat};
\addplot file {./figs/empirical/perfect/successrate/brazzers_1e6.dat};

\path[name path=B] (axis cs:(9 * 1e7,0) -- (axis cs:(9 * 1e7,1);
\path[name path=A1] (axis cs:(3 *1e5,0) -- (axis cs:(3 * 1e5,1);
\tikzfillbetween[of=A1 and B, on layer=main]{red, opacity=0.2};

\path[name path=A] (axis cs:(2 *1e5,0) -- (axis cs:(2 * 1e5,1);
\tikzfillbetween[of=A and B, on layer=main]{yellow, opacity=0.2};

\node at (axis cs:5*1e7,0.5) {\rotatebox{90}{\tiny uncertain region}};

\end{semilogxaxis}
\end{tikzpicture}
\label{fig:empiricalbrazzers}
}
\hfill
\subfloat[Clixsense]{
\begin{tikzpicture}[scale=0.6]
\begin{semilogxaxis}[ymin=0]

\addplot+[stack plots=y] file {./figs/empirical/perfect/successrate/clixsense.dat};
\addplot file {./figs/empirical/perfect/successrate/clixsense11by3.dat};
\addplot file {./figs/empirical/imperfect/successrate/clixsense11by3_test.dat};
\addplot+[stack plots=y, stack dir=minus] file {./figs/empirical/perfect/successrate/clixsense11by3.dat};

\addplot file {./figs/empirical/perfect/successrate/clixsense_1e5.dat};
\addplot file {./figs/empirical/perfect/successrate/clixsense_1e6.dat};

\path[name path=A] (axis cs:(4 * 1e5,0) -- (axis cs:(4 * 1e5,1);
\path[name path=B] (axis cs:(9 * 1e7,0) -- (axis cs:(9 * 1e7,1);
\tikzfillbetween[of=A and B, on layer=main]{yellow, opacity=0.2};

\path[name path=A1] (axis cs:(8 * 1e5,0) -- (axis cs:(8 * 1e5,1);
\tikzfillbetween[of=A1 and B, on layer=main]{red, opacity=0.2};
\node at (axis cs:5*1e7,0.5) {\rotatebox{90}{\tiny uncertain region}};

\end{semilogxaxis}
\end{tikzpicture}
\label{fig:empiricalclixsense}
}

\subfloat[CSDN]{
\begin{tikzpicture}[scale=0.6]
\begin{semilogxaxis}[ymin=0]
\addplot+[stack plots=y] file {./figs/empirical/perfect/successrate/csdn.dat};
\addplot file {./figs/empirical/perfect/successrate/csdn11by3.dat};
\addplot file {./figs/empirical/imperfect/successrate/csdn11by3_test.dat};
\addplot+[stack plots=y, stack dir=minus] file {./figs/empirical/perfect/successrate/csdn11by3.dat};

\addplot file {./figs/empirical/perfect/successrate/csdn_1e5.dat};
\addplot file {./figs/empirical/perfect/successrate/csdn_1e6.dat};

\path[name path=A] (axis cs:(2* 1e6,0) -- (axis cs:(2* 1e6,1);
\path[name path=B] (axis cs:(9 * 1e7,0) -- (axis cs:(9 * 1e7,1);
\tikzfillbetween[of=A and B, on layer=main]{red, opacity=0.2};

\path[name path=A1] (axis cs:(1* 1e6,0) -- (axis cs:(1* 1e6,1);
\tikzfillbetween[of=A1 and B, on layer=main]{yellow, opacity=0.2};
\node at (axis cs:5*1e7,0.5) {\rotatebox{90}{\tiny uncertain region}};

\end{semilogxaxis}
\end{tikzpicture}
\label{fig:empiricalcsdn}
}
\hfill
\subfloat[Linkedin]{
\begin{tikzpicture}[scale=0.6]
\begin{semilogxaxis}[ymin=0]
\addplot+[stack plots=y] file {./figs/empirical/perfect/successrate/linkedin.dat};
\addplot file {./figs/empirical/perfect/successrate/linkedin11by3.dat};
\addplot file {./figs/empirical/imperfect/successrate/linkedin11by3_test.dat};
\addplot+[stack plots=y, stack dir=minus] file {./figs/empirical/perfect/successrate/linkedin11by3.dat};

\addplot file {./figs/empirical/perfect/successrate/linkedin_1e5.dat};
\addplot file {./figs/empirical/perfect/successrate/linkedin_1e6.dat};

\path[name path=A] (axis cs:(3* 1e7,0) -- (axis cs:(3* 1e7,1);
\path[name path=B] (axis cs:(9 * 1e7,0) -- (axis cs:(9 * 1e7,1);
\tikzfillbetween[of=A and B, on layer=main]{red, opacity=0.2};

\path[name path=A1] (axis cs:(2* 1e7,0) -- (axis cs:(2* 1e7,1);
\tikzfillbetween[of=A1 and B, on layer=main]{yellow, opacity=0.2};

\node at (axis cs:5*1e7,0.5) {\rotatebox{90}{\tiny uncertain region}};

\end{semilogxaxis}
\end{tikzpicture}
\label{fig:empiricallinkedin}
}
\hfill
\subfloat[Neopets]{
\begin{tikzpicture}[scale=0.6]
\begin{semilogxaxis}[ymin=0]

\addplot+[stack plots=y] file {./figs/empirical/perfect/successrate/neopets.dat};
\addplot file {./figs/empirical/perfect/successrate/neopets11by3.dat};
\addplot file {./figs/empirical/imperfect/successrate/neopets11by3_test.dat};
\addplot+[stack plots=y, stack dir=minus] file {./figs/empirical/perfect/successrate/neopets11by3.dat};

\addplot file {./figs/empirical/perfect/successrate/neopets_1e5.dat};
\addplot file {./figs/empirical/perfect/successrate/neopets_1e6.dat};

\path[name path=A] (axis cs:(1.5* 1e7,0) -- (axis cs:(1.5* 1e7,1);
\path[name path=B] (axis cs:(9 * 1e7,0) -- (axis cs:(9 * 1e7,1);
\tikzfillbetween[of=A and B, on layer=main]{red, opacity=0.2};

\path[name path=A1] (axis cs:(5* 1e6,0) -- (axis cs:(5* 1e6,1);
\tikzfillbetween[of=A1 and B, on layer=main]{yellow, opacity=0.2};

\node at (axis cs:5*1e7,0.5) {\rotatebox{90}{\tiny uncertain region}};

\end{semilogxaxis}
\end{tikzpicture}
\label{fig:empiricalneopets}
}

\caption{Adversary Success Rate vs $v/C_{max}$ for Empirical Distributions} {\par \small the red (resp. yellow) shaded areas denote unconfident regions where the the empirical distribution might diverges from the real distribution $E \geq 0.1$ (resp. $E \geq 0.01$).} 
\label{fig:empirical}
\vspace{-0.4cm}
\end{figure*}

%% file: figs/unlucky.tex
\tikzUnlucky
\begin{figure*}[ht]\centering
\subfloat[bfield, brazzers, clixsense]{
\begin{tikzpicture}[scale=0.6]
\begin{semilogxaxis}[enlargelimits=false]

\addplot+[only marks] file{./figs/empirical/unlucky/successrate/bfield11by3.dat};
\addlegendentry{bfield}
\addplot+[only marks] file{./figs/empirical/unlucky/successrate/brazzers11by3.dat};
\addlegendentry{brazzers}
\addplot+[only marks] file{./figs/empirical/unlucky/successrate/clixsense11by3.dat};
\addlegendentry{clixsense}

\end{semilogxaxis}
\end{tikzpicture}

}
\hfill
\subfloat[csdn, LinkedIn, neopets ]{
\begin{tikzpicture}[scale=0.6]
\begin{semilogxaxis}[ymin=0]

\addplot+[only marks] file{./figs/empirical/unlucky/successrate/csdn11by3.dat};
\addlegendentry{csdn}
\addplot+[only marks] file{./figs/empirical/unlucky/successrate/linkedin11by3.dat};
\addlegendentry{linkedin}
\addplot+[only marks] file{./figs/empirical/unlucky/successrate/neopets11by3.dat};
\addlegendentry{neopets}

\end{semilogxaxis}
\end{tikzpicture}
}
\hfill
\subfloat[RockYou, 000webhost, Yahoo!]{
\begin{tikzpicture}[scale=0.6]
\begin{semilogxaxis}[ymin=0]

\addplot+[only marks] file{./figs/empirical/unlucky/successrate/rockyou11by3.dat};
\addlegendentry{rockyou}
\addplot+[only marks] file{./figs/empirical/unlucky/successrate/000webhost11by3.dat};
\addlegendentry{000webhost}
\addplot+[only marks] file{./figs/empirical/unlucky/successrate/yahoo11by3.dat};
\addlegendentry{yahoo}

\end{semilogxaxis}
\end{tikzpicture}
}
\caption{Proportion of Unlucky Users for Various Datasets ($E\left[X_u\right]$)}
\label{fig:unlucky}
\vspace{-0.4cm}
\end{figure*}
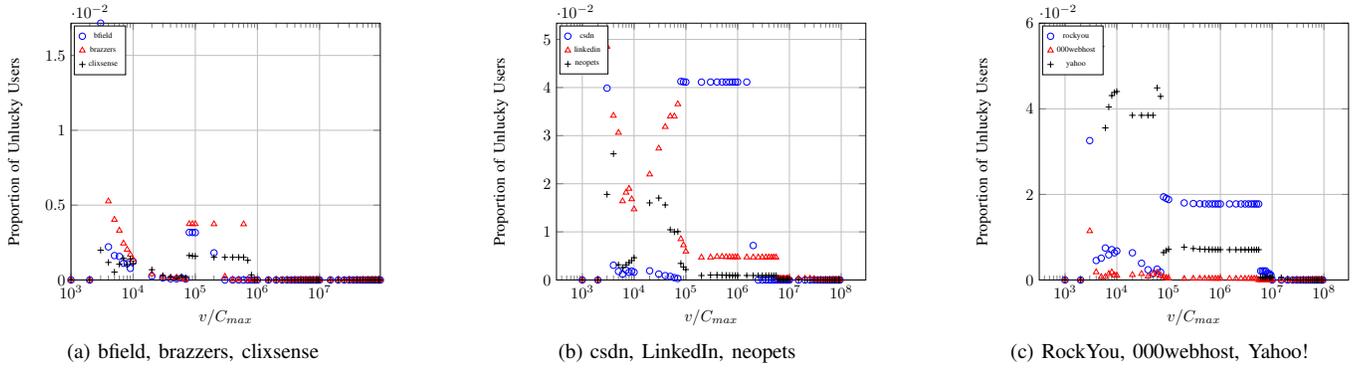

%% file: figs/monte.tex
\tikzMonte
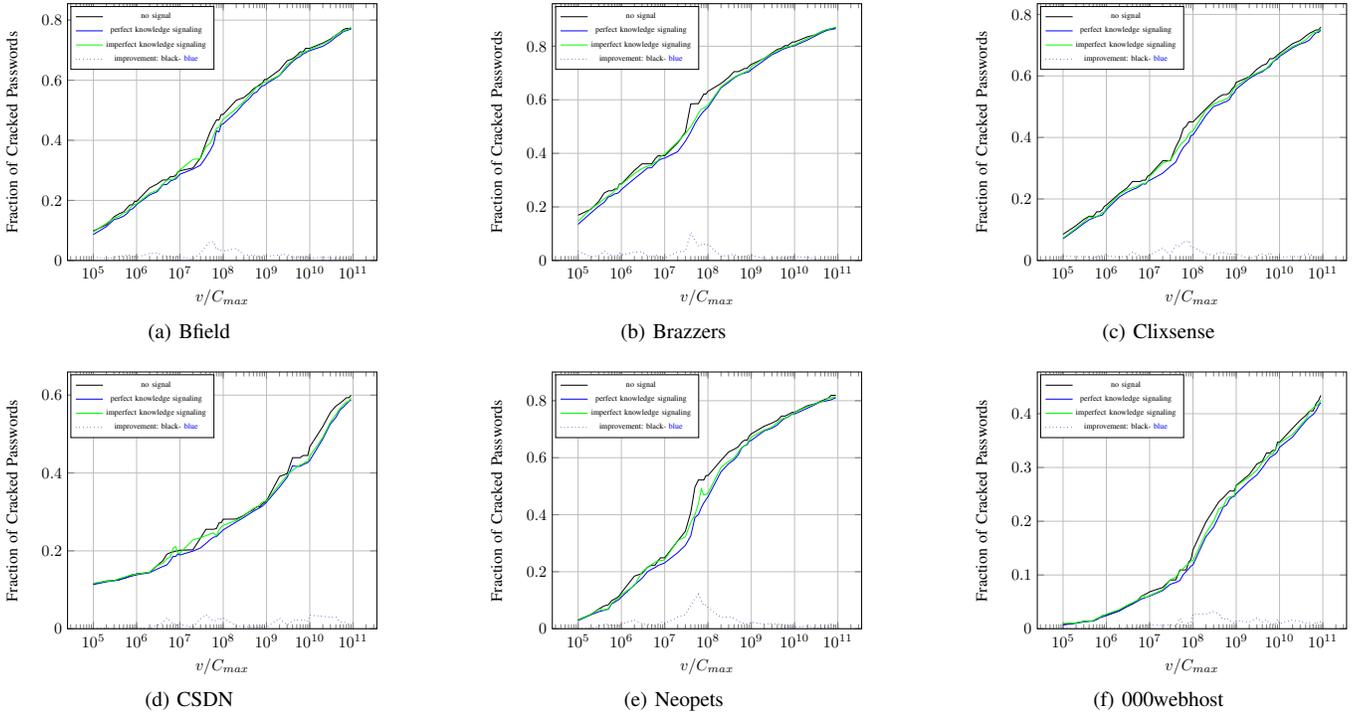
\begin{figure*}[ht]\centering
\subfloat[Bfield]{
\begin{tikzpicture}[scale=0.6]
\begin{semilogxaxis}[ymin=0]

\addplot+[stack plots=y] file {./figs/montecarlo/perfect/successrate/bfield.dat};
\addplot file {./figs/montecarlo/perfect/successrate/bfield11by3.dat};
\addplot file {./figs/montecarlo/imperfect/successrate/bfield11by3_test.dat};
\addplot+[stack plots=y, stack dir=minus] file {./figs/montecarlo/perfect/successrate/bfield11by3.dat};

\end{semilogxaxis}
\end{tikzpicture}
\label{fig:montebfield}
}
\hfill
\subfloat[Brazzers]{
\begin{tikzpicture}[scale=0.6]
\begin{semilogxaxis}[ymin=0]

\addplot+[stack plots=y]  file {./figs/montecarlo/perfect/successrate/brazzers.dat};
\addplot file {./figs/montecarlo/perfect/successrate/brazzers11by3.dat};
\addplot file {./figs/montecarlo/imperfect/successrate/brazzers11by3_test.dat};
\addplot+[stack plots=y, stack dir=minus] file {./figs/montecarlo/perfect/successrate/brazzers11by3.dat};

\end{semilogxaxis}
\end{tikzpicture}
\label{fig:montebrazzers}
}
\hfill
\subfloat[Clixsense]{
\begin{tikzpicture}[scale=0.6]
\begin{semilogxaxis}[ymin=0]
\addplot+[stack plots=y]  file {./figs/montecarlo/perfect/successrate/clixsense.dat};
\addplot file {./figs/montecarlo/perfect/successrate/clixsense11by3.dat};
\addplot file {./figs/montecarlo/imperfect/successrate/clixsense11by3_test.dat};
\addplot+[stack plots=y, stack dir=minus] file {./figs/montecarlo/perfect/successrate/clixsense11by3.dat};

\end{semilogxaxis}
\end{tikzpicture}
\label{fig:monteclixsense}
}

\subfloat[CSDN]{
\begin{tikzpicture}[scale=0.6]
\begin{semilogxaxis}[ymin=0]
\addplot+[stack plots=y]  file {./figs/montecarlo/perfect/successrate/csdn.dat};
\addplot file {./figs/montecarlo/perfect/successrate/csdn11by3.dat};
\addplot file {./figs/montecarlo/imperfect/successrate/csdn11by3_test.dat};
\addplot+[stack plots=y, stack dir=minus] file {./figs/montecarlo/perfect/successrate/csdn11by3.dat};

\end{semilogxaxis}
\end{tikzpicture}
\label{fig:montecsdn}
}
\hfill
\subfloat[Neopets]{
\begin{tikzpicture}[scale=0.6]
\begin{semilogxaxis}[ymin=0]
\addplot+[stack plots=y]  file {./figs/montecarlo/perfect/successrate/neopets.dat};
\addplot file {./figs/montecarlo/perfect/successrate/neopets11by3.dat};
\addplot file {./figs/montecarlo/imperfect/successrate/neopets11by3_test.dat};
\addplot+[stack plots=y, stack dir=minus] file {./figs/montecarlo/perfect/successrate/neopets11by3.dat};

\end{semilogxaxis}
\end{tikzpicture}
\label{fig:monteneopets}
}
\hfill
\subfloat[000webhost]{
\begin{tikzpicture}[scale=0.6]
\begin{semilogxaxis}[ymin=0]
\addplot+[stack plots=y]  file {./figs/montecarlo/perfect/successrate/000webhost.dat};
\addplot file {./figs/montecarlo/perfect/successrate/000webhost11by3.dat};
\addplot file {./figs/montecarlo/imperfect/successrate/000webhost11by3_test.dat};
\addplot+[stack plots=y, stack dir=minus] file {./figs/montecarlo/perfect/successrate/000webhost11by3.dat};

\end{semilogxaxis}
\end{tikzpicture}
\label{fig:montewebhost}
}
\caption{Adversary Success Rate vs $v/k$ for Monte Carlo Distributions}
\label{fig:monte}
\vspace{-0.4cm}
\end{figure*}

%% file: figs/online.tex
\tikzOnline
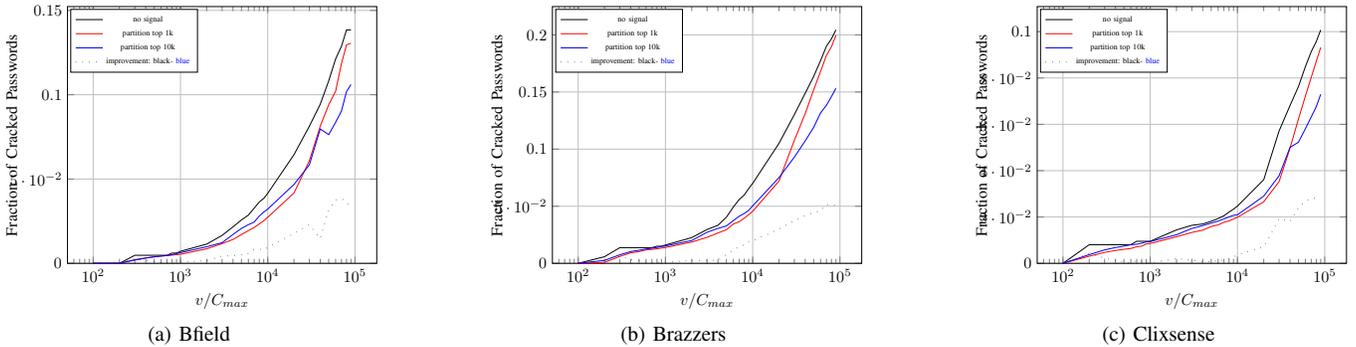
\begin{figure*}[ht]\centering
\subfloat[Bfield]{
\begin{tikzpicture}[scale=0.6]
\begin{semilogxaxis}[ymin=0]

\addplot+[stack plots=y] file {./figs/empirical/online/successrate/bfield.dat};
\addplot file {./figs/empirical/online/successrate/bfieldtop1k.dat};
\addplot file {./figs/empirical/online/successrate/bfieldtop10k.dat};
\addplot+[stack plots=y, stack dir=minus] file {./figs/empirical/online/successrate/bfieldtop10k.dat};
\end{semilogxaxis}
\end{tikzpicture}
\label{fig:onlinebfield}
}
\hfill
\subfloat[Brazzers]{
\begin{tikzpicture}[scale=0.6]
\begin{semilogxaxis}[ymin=0]
\addplot+[stack plots=y] file {./figs/empirical/online/successrate/brazzers.dat};
\addplot file {./figs/empirical/online/successrate/brazzerstop1k.dat};
\addplot file {./figs/empirical/online/successrate/brazzerstop10k.dat};
 \addplot+[stack plots=y, stack dir=minus] file {./figs/empirical/online/successrate/brazzerstop10k.dat};
\end{semilogxaxis}
\end{tikzpicture}
\label{fig:onlinebrazzers}
}
\hfill
\subfloat[Clixsense]{
\begin{tikzpicture}[scale=0.6]
\begin{semilogxaxis}[ymin=0]
\addplot+[stack plots=y] file {./figs/empirical/online/successrate/clixsense.dat};
\addplot file {./figs/empirical/online/successrate/clixsensetop1k.dat};
\addplot file {./figs/empirical/online/successrate/clixsensetop10k.dat};
\addplot+[stack plots=y, stack dir=minus] file {./figs/empirical/online/successrate/clixsensetop10k.dat};
\end{semilogxaxis}
\end{tikzpicture}
\label{fig:onlineclixsense}
}

\caption{Adversary Success Rate vs $v/C_{max}$ in Defense of Online Attacks} 
\label{fig:online}
\vspace{-0.4cm}
\end{figure*}

%% file: sections/conclusion.tex

We introduced password strength signaling as a novel, yet counter-intuitive defense against rational password attackers. We use Stackelberg game to model the interaction between the defender and attacker, and present an algorithm for the server to optimize its signaling matrix. We ran experiments to empirically evaluate the effectiveness of information signaling on 9  password datasets. When testing on the empirical  (resp. Monte Carlo) password distribution distribution we find that information signaling reduces the number of passwords that would have been cracked by up to $8\%$ (resp. $12\%$). Additionally, we find that information signaling can help to dissuade an online attacker by saving $5\%$ of all user accounts. We view our positive experimental results as a proof of concept which motivates further exploration of password strength signaling.


%% file: sections/AppendixCompressedDistribution.tex
\section*{Compressed Password Distributions}\label{appendix:es}
Given a distribution over $N$ passwords with corresponding probabilities $p_1\geq p_2 \geq \ldots p_N$ we can often encode the probabilities in compressed form by grouping passwords with equal probability. Viewed in this way we can encode the distribution as a sequence of $N' \leq N$ tuples $(p_1,c_1),\ldots,(p_{N'},c_N')$ where $p_1 > p_2 > \ldots > p_{N'}$ and $c_i$ denotes the number of passwords with probability exactly $p_i$ in the distribution. In all of the distributions we analyze we have $N' \ll N$ e.g., for the RockYou empirical we have $N \geq 3.2 \times 10^7$ while $N' \leq 2.3 \times 10^3$. Thus, it is desirable to ensure that our optimization algorithms scale with $N'$ instead of $N$. Similar observations were used by Blocki and Datta~\cite{CSF:BloDat16} and Bai and Blocki~\cite{DAHash}.

Bai and Blocki~\cite[Lemma 1]{DAHash} showed that a rational adversary never {\em splits} equivalence sets i.e., if $\Pr[pw_i] = \Pr[pw_j]$ then the attacker will either check both passwords or neither. Thus, a rational attacker's optimal strategy will be to check the $B^*$ most popular passwords where $B^*$ is guaranteed to be in the set $\{0, c_1, c_1+c_2,\ldots, \sum_{i=1}^{N'} c_i\}$. When the distribution is compact this substantially narrows down the search space in comparison to a brute-force search over all possible choices of $B^* \in [N]$.

We observe that if the original password distribution has a compact representation $(p_1,c_1),\ldots,(p_{N'},c_N')$ with $N'$ equivalence sets then, after observing the password signaling $y \in [b]$, the posterior distribution has a compact representation with {\em at most} $aN'$ equivalence sets where the dimension of the signaling matrix is $a \times b$ i.e., $\mathbf{S} \in \mathbb{R}^{a \times b}$. To see this notice that we can partition all passwords $pw$ in equivalence set $i$ into $a$ groups based on the value $\mathsf{getStrength}(pw) \in [a]$. If $\Pr[pw]=\Pr[pw']$ and $\mathsf{getStrength}(pw) = \mathsf{getStrength(pw)}$ then the posterior probabilities are also equal i.e., $\Pr[pw| y] = \Pr[pw' | y]$. 

Thus, given a signaling matrix $\mathbf{S}$ and a signal $y\in[b]$ and we can compute the adversaries optimal response $(\pi_b^*, B_y^*)$ to the signal $y$ by 1) computing the compact representation of the posterior distribution, and 2) checking all $aN'$ possible values of $B_y^*$  to find the budget that maximizes the attacker's expected utility conditioning on the signal $y$. After pre-processing the original dataset the first step only requires time $O(aN' \log aN')$ for each new signaling matrix $\mathbf{S}$ and $y \in [b]$.

%% file: figs/emprical_extra.tex
\tikzEmpirical
\begin{figure*}[ht]\centering
\subfloat[Rockyou]{
\begin{tikzpicture}[scale=0.65]
\begin{semilogxaxis}[ymin=0]
\addplot+[stack plots=y] file {./figs/empirical/perfect/successrate/rockyou.dat};
\addplot file {./figs/empirical/perfect/successrate/rockyou11by3.dat};
\addplot file {./figs/empirical/imperfect/successrate/rockyou11by3_test.dat};
\addplot+[stack plots=y, stack dir=minus] file {./figs/empirical/perfect/successrate/rockyou11by3.dat};

\addplot file {./figs/empirical/perfect/successrate/rockyou_1e5.dat};
\addplot file {./figs/empirical/perfect/successrate/rockyou_1e6.dat};

\path[name path=A] (axis cs:(7 * 1e6,0) -- (axis cs:(7 * 1e6,1);
\path[name path=B] (axis cs:(9 * 1e7,0) -- (axis cs:(9 * 1e7,1);
\tikzfillbetween[of=A and B, on layer=main]{red, opacity=0.2};

\path[name path=A1] (axis cs:(3 * 1e6,0) -- (axis cs:(3 * 1e6,1);
\tikzfillbetween[of=A1 and B, on layer=main]{yellow, opacity=0.2};

\node at (axis cs:5*1e7,0.5) {\rotatebox{90}{\tiny uncertain region}};

\end{semilogxaxis}
\end{tikzpicture}
\label{fig:empiricalrockyou}
}
\hfill
\subfloat[000webhost]{
\begin{tikzpicture}[scale=0.65]
\begin{semilogxaxis}[ymin=0]
\addplot+[stack plots=y] file {./figs/empirical/perfect/successrate/000webhost.dat};
\addplot file {./figs/empirical/perfect/successrate/000webhost11by3.dat};
\addplot file {./figs/empirical/imperfect/successrate/000webhost11by3_test.dat};
\addplot+[stack plots=y, stack dir=minus] file {./figs/empirical/perfect/successrate/000webhost11by3.dat};

\addplot file {./figs/empirical/perfect/successrate/000webhost_1e5.dat};
\addplot file {./figs/empirical/perfect/successrate/000webhost_1e6.dat};

\path[name path=A] (axis cs: 5* 1e6,0) -- (axis cs: 5* 1e6,1);
\path[name path=B] (axis cs: 9 * 1e7,0) -- (axis cs:9 * 1e7,1);
\tikzfillbetween[of=A and B, on layer=main]{red, opacity=0.2};

\path[name path=A1] (axis cs: 2* 1e6,0) -- (axis cs: 2* 1e6,1);
\tikzfillbetween[of=A1 and B, on layer=main]{yellow, opacity=0.2};

\node at (axis cs:5*1e7,0.5) {\rotatebox{90}{\tiny uncertain region}};

\end{semilogxaxis}
\end{tikzpicture}
\label{fig:empiricalwebhost}
}
\hfill
\subfloat[Yahoo]{
\begin{tikzpicture}[scale=0.65]
\begin{semilogxaxis}[ymin=0]
\addplot+[stack plots=y] file {./figs/empirical/perfect/successrate/yahoo.dat};
\addplot file {./figs/empirical/perfect/successrate/yahoo11by3.dat};
\addplot file {./figs/empirical/imperfect/successrate/yahoo11by3_test.dat};
\addplot+[stack plots=y, stack dir=minus] file {./figs/empirical/perfect/successrate/yahoo11by3.dat};

\addplot file {./figs/empirical/perfect/successrate/yahoo_1e5.dat};
\addplot file {./figs/empirical/perfect/successrate/yahoo_1e6.dat};

\path[name path=A] (axis cs:(2* 1e7,0) -- (axis cs:(2* 1e7,1);
\path[name path=B] (axis cs:(9 * 1e7,0) -- (axis cs:(9 * 1e7,1);
\tikzfillbetween[of=A and B, on layer=main]{red, opacity=0.2};

\path[name path=A1] (axis cs:(7* 1e6,0) -- (axis cs:(7* 1e6,1);
\tikzfillbetween[of=A1 and B, on layer=main]{yellow, opacity=0.2};

\node at (axis cs:5*1e7,0.5) {\rotatebox{90}{\tiny uncertain region}};

\end{semilogxaxis}
\end{tikzpicture}
\label{fig:empiricalYahoo}
}
\caption{Adversary Success Rate vs $v/C_{max}$ for Empirical Distributions} {\par \small the red (resp. yellow) shaded areas denote unconfident regions where the the empirical distribution might diverges from the real distribution $E \geq 0.1$ (resp. $E \geq 0.01$).} 
\label{fig:empiricalExtra}
\vspace{-0.4cm}
\end{figure*}
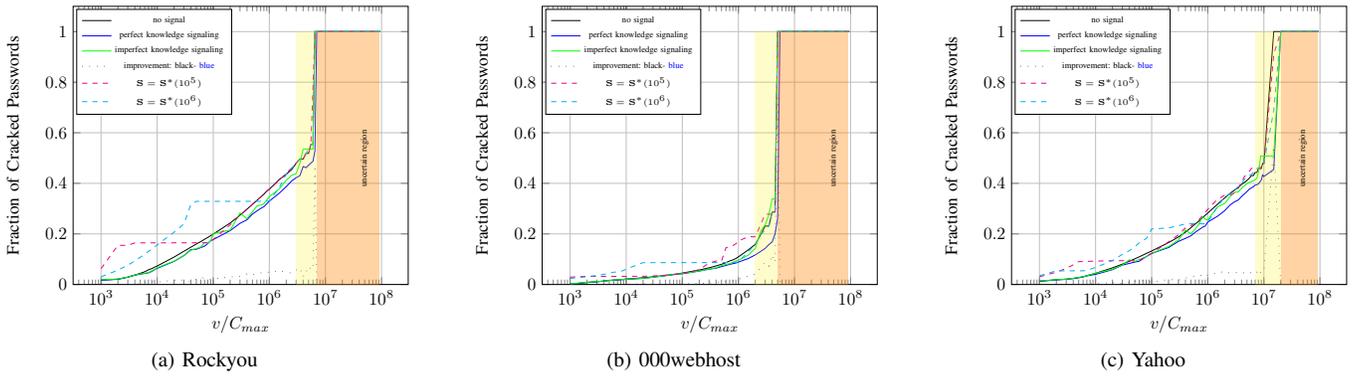

%% file: figs/online_extra.tex
\tikzOnline
\begin{figure*}[ht]\centering

\subfloat[Rockyou]{
\begin{tikzpicture}[scale=0.65]
\begin{semilogxaxis}[ymin=0]
\addplot+[stack plots=y] file {./figs/empirical/online/successrate/rockyou.dat};
\addplot file {./figs/empirical/online/successrate/rockyoutop1k.dat};
\addplot file {./figs/empirical/online/successrate/rockyoutop10k.dat};
\addplot+[stack plots=y, stack dir=minus] file {./figs/empirical/online/successrate/rockyoutop10k.dat};

\end{semilogxaxis}
\end{tikzpicture}
\label{fig:onlinerockyou}
}
\hfill
\subfloat[000webhost]{
\begin{tikzpicture}[scale=0.65]
\begin{semilogxaxis}[ymin=0]
\addplot+[stack plots=y] file {./figs/empirical/online/successrate/000webhost.dat};
\addplot file {./figs/empirical/online/successrate/000webhosttop1k.dat};
\addplot file {./figs/empirical/online/successrate/000webhosttop10k.dat};
\addplot+[stack plots=y, stack dir=minus] file {./figs/empirical/online/successrate/000webhosttop10k.dat};

\end{semilogxaxis}
\end{tikzpicture}
\label{fig:onlinewebhost}
}
\hfill
\subfloat[Yahoo]{
\begin{tikzpicture}[scale=0.65]
\begin{semilogxaxis}[ymin=0]
\addplot+[stack plots=y] file {./figs/empirical/online/successrate/yahoo.dat};
\addplot file {./figs/empirical/online/successrate/yahootop1k.dat};
\addplot file {./figs/empirical/online/successrate/yahootop10k.dat};
\addplot+[stack plots=y, stack dir=minus] file {./figs/empirical/online/successrate/yahootop10k.dat};

\end{semilogxaxis}
\end{tikzpicture}
\label{fig:onlineYahoo}
}

\subfloat[CSDN]{
\begin{tikzpicture}[scale=0.65]
\begin{semilogxaxis}[ymin=0]
\addplot+[stack plots=y] file {./figs/empirical/online/successrate/csdn.dat};
\addplot file {./figs/empirical/online/successrate/csdntop1k.dat};
\addplot file {./figs/empirical/online/successrate/csdntop10k.dat};
\addplot+[stack plots=y, stack dir=minus] file {./figs/empirical/online/successrate/csdntop10k.dat};

\end{semilogxaxis}
\end{tikzpicture}
\label{fig:onlinecsdn}
}
\hfill
\subfloat[Linkedin]{
\begin{tikzpicture}[scale=0.65]
\begin{semilogxaxis}[ymin=0]
\addplot+[stack plots=y] file {./figs/empirical/online/successrate/linkedin.dat};
\addplot file {./figs/empirical/online/successrate/linkedintop1k.dat};
\addplot file {./figs/empirical/online/successrate/linkedintop10k.dat};
\addplot+[stack plots=y, stack dir=minus] file {./figs/empirical/online/successrate/linkedintop10k.dat};

\end{semilogxaxis}
\end{tikzpicture}
\label{fig:onlinelinkedin}
}
\hfill
\subfloat[Neopets]{
\begin{tikzpicture}[scale=0.65]
\begin{semilogxaxis}[ymin=0]

\addplot+[stack plots=y] file {./figs/empirical/online/successrate/neopets.dat};
\addplot file {./figs/empirical/online/successrate/neopetstop1k.dat};
\addplot file {./figs/empirical/online/successrate/neopetstop10k.dat};
\addplot+[stack plots=y, stack dir=minus] file {./figs/empirical/online/successrate/neopetstop10k.dat};

\end{semilogxaxis}
\end{tikzpicture}
\label{fig:onlineneopets}
}

\caption{Adversary Success Rate vs $v/C_{max}$ in Defense of Online Attacks} 
\label{fig:onlineExtra}
\vspace{-0.4cm}
\end{figure*}
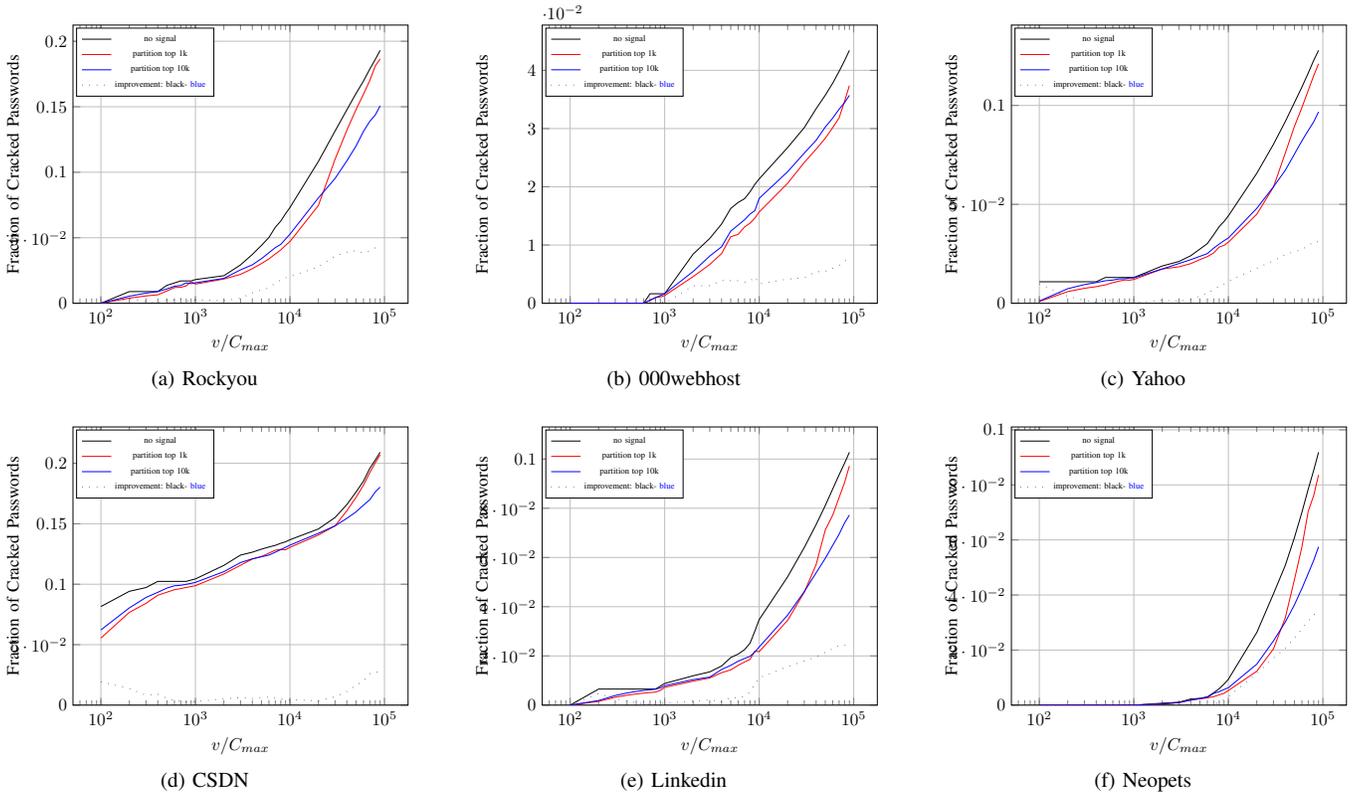